\documentclass[aps,prd,twocolumn,superscriptaddress,nofootinbib]{revtex4-2}

\usepackage{graphicx}
\usepackage{amsmath}
\usepackage{amssymb}
\usepackage{hyperref}
\usepackage{xcolor}
\definecolor{darkgreen}{RGB}{0,120,0} 
\usepackage{siunitx}
\usepackage{mathalfa}
\usepackage[normalem]{ulem} 
\usepackage{cancel}
\usepackage[normalem]{ulem}

\def\beq{\begin{equation}}
\def\eeq{\end{equation}}

\newcommand{\rs}{r}
\newcommand{\Rs}{R}

\newcommand{\fw}{f_{\mathrm{wall}}}

\begin{document}
\begin{flushright}
NORDITA 2025-061
\end{flushright}

\title{String-induced vacuum decay and its gravitational wave signatures}

\author{Aleksandr Chatrchyan}
\email{aleksandr.chatrchyan@su.se}
\affiliation{
The Oskar Klein Centre for Cosmoparticle Physics,
Department of Physics, Stockholm University, AlbaNova, 10691 Stockholm, Sweden}
\affiliation{
Nordita, KTH Royal Institute of Technology and Stockholm University\\
Hannes Alfv\'ens v\"ag 12, SE-106 91 Stockholm, Sweden}
\author{Florian Niedermann}
\email{florian.niedermann@su.se}
\affiliation{
Nordita, KTH Royal Institute of Technology and Stockholm University\\
Hannes Alfv\'ens v\"ag 12, SE-106 91 Stockholm, Sweden}
\author{Phoebe Richman-Taylor}
\email{pmr2@hi.is}
\affiliation{Center for Astrophysics and Cosmology, Science Institute,
University of Iceland, Dunhagi 5, 107 Reykjavik, Iceland}

\date{\today}

\begin{abstract}
False vacuum decay typically proceeds via the nucleation of spherical bubbles of true vacuum, described by $O(4)$ symmetric field configurations in Euclidean time. In this work, we investigate how the presence of cosmic strings can catalyze the decay process. To this end, we consider a complex scalar field charged under a global or local $U(1)$ symmetry. Assuming a non-trivial vacuum manifold, realizable for example in a simple sextic potential, we derive relativistic bounce solutions with $O(2) \times O(2)$ symmetry, corresponding to elongated bubbles seeded by a cosmic string of the same scalar field. Building up on earlier results in the literature, we identify the region of parameter space where vacuum decay predominantly proceeds via this alternative channel, thereby providing an explicit mechanism for the quantum decay of cosmic strings. Finally, we present an initial discussion of the gravitational wave signal associated with this type of vacuum decay and its possible connection to the recently observed stochastic signal in pulsar timing arrays.
\end{abstract}

\maketitle

\tableofcontents

\section{Introduction}

\begin{figure}[b!]
  \centering
  \includegraphics[width=0.75\linewidth]{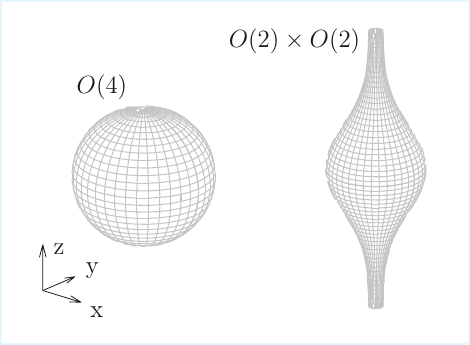}
  \caption{Schematic illustration of the Coleman $O(4)$ bounce (left), and the $O(2)\times O(2)$ bounce in the presence of a cosmic string (right).}
  \label{fig:illustration}
\end{figure}

In the standard treatment of vacuum metastability, the decay of the false vacuum is described in terms of the \textit{bounce}, which is an $O(4)$ symmetric solution of the Euclidean equations of motion~\cite{Coleman:1977py,Callan:1977pt} (left illustration in Fig.~\ref{fig:illustration}). The initial state is typically taken to be the Lorentz-invariant Minkowski vacuum, at least locally. It is therefore natural to ask how the decay process is modified when the initial state \textit{breaks} Lorentz symmetry. A simple and well-motivated example is a straight cosmic string, corresponding to an $O(2)$ symmetric configuration of a complex scalar field. This configuration is classically stable due to a nontrivial winding of the field's phase, which topologically protects it from unwinding into the trivial Minkowski vacuum. 

In this work, we show that for a general class of scalar-field potentials, consistent with an effective field theory framework, the system admits $O(2) \times O(2)$ symmetric bounce solutions describing the sub-barrier tunneling of a metastable cosmic string. In this process, the string transitions to a new state in which its core becomes unstable and expands to infinity (right illustration in Fig.~\ref{fig:illustration}), leading to the disappearance of the string and the restoration of full Lorentz symmetry.

As a minimal setting, we consider a complex scalar field with a sextic potential charged under a global or local $U(1)$ symmetry. The potential features a spontaneously broken false vacuum separated by a barrier from the symmetric true vacuum. In the first part of this work, we demonstrate that this system supports $O(2) \times O(2)$ symmetric solutions to its Euclidean equations of motion, compute the associated bounce action, and use it to estimate the string-induced decay rate of the symmetry-broken false vacuum to the symmetric true vacuum.

The string-induced decay of the false vacuum in a first-order phase transition has been investigated in several earlier works. 
In Ref.~\cite{Dasgupta_1997}, Dasgupta discussed $O(2)\times O(2)$ symmetric bounce solutions and proved their existence within the Abelian Higgs model, albeit without providing an explicit calculation of the bounce action or the corresponding tunneling rate. More recently, Lee \textit{et al.}~\cite{Lee:2013zca}, building on earlier work in 2+1 dimensions~\cite{Lee:2013ega}, computed the bounce action in the thin-wall limit by mapping the field-theoretic problem to a quantum-mechanical one. Their treatment, however, does not fully explore the thin-wall parameter regime and neglects relativistic corrections to the bounce dynamics, limiting the range of applicability of their results to small wall velocities. 

Another form of string metastability, distinct from the mechanism considered here, was first studied by Vilenkin in Ref.~\cite{Vilenkin:1982hm}, where monopole–antimonopole pairs nucleate along the string core, fragmenting it into finite segments. In contrast, our scenario involves the nucleation of true-vacuum bubbles along the string core, which subsequently expand and convert the entire space to the true vacuum, thereby erasing the string network and restoring the symmetry. 
Finally, as discussed in~\cite{Steinhardt:1981ec, Steinhardt:1981mm} (and~\cite{Jensen:1982jv}) topological defects can also catalyze a \textit{classical} decay of the false vacuum when monopoles and other defects “dissociate” (see also~\cite{Yajnik:1986tg, Yajnik:1986wq} for the dissociation of cosmic strings in the context of an $SO(10)$ GUT model). In~\cite{Blasi:2024mtc} high-temperature thermal tunneling induced by axionic (global) cosmic strings was studied in the context of the electroweak phase transition.  
Induced bubble nucleation was also studied in the context of other topological defects, such as monopoles~\cite{Kumar:2010mv, Agrawal:2022hnf}, domain walls~\cite{Blasi:2022woz, Blasi:2023rqi, Agrawal:2023cgp}, non-topological solitons such as Q-balls~\cite{Metaxas:2000qf}, oscillons~\cite{Gleiser:2007ts} and even black holes~\cite{Moss:1984zf, Hiscock:1987hn, Berezin:1987ea, Arnold:1989cq, Gregory:2013hja, Mukaida:2017bgd, Shkerin:2021zbf, Jinno:2023vnr} (see also Refs.~\cite{Billam:2018pvp, Billam:2022ykl, Jenkins:2025szl} for a discussion of how these effects can be probed in cold atomic systems).

Our tunneling analysis follows a similar strategy to Ref.~\cite{Lee:2013zca}, where the field-theoretic problem is mapped to the problem of describing the dynamics of the string wall. However, our treatment improves upon that work in several crucial ways.
First, we provide a fully relativistic treatment of the bubble-wall dynamics, including the effects of Lorentz contraction that are crucial for accurately describing the post-phase transition dynamics of the string wall and gaining higher precision in calculating the bounce action.
Second, we perform a comprehensive exploration of the parameter space for both global and local strings in the thin-wall limit, deriving explicit constraints that delineate the regime in which string-induced tunneling is enhanced relative to the standard (string-less) vacuum decay. This includes explicit semi-analytic expression for the bounce action and the bounce radius. 

Throughout the analysis, we place particular emphasis on assessing the validity of the thin-wall approximation, identifying the region where it provides a quantitatively reliable description.
As the main result, we find that the string-induced channel dominates false vacuum decay in a broad parameter range, rendering the standard decay through $O(4)$ symmetric Coleman bubbles irrelevant within this regime.

Having established the phenomenological relevance of the mechanism, we turn to its cosmological implications for gravitational waves (GW). The crucial observation is that a single $O(2)\times O(2)$ bubble exhibits a non-vanishing and time-dependent quadrupole moment $Q(t)$. As a result, an isolated expanding bubble emits GWs, in contrast to a spherically symmetric one. We compute the quadrupole moment and numerically infer its scaling relation 
with the bubble radius $R(t)$.  We then compare the strength of the signal with the 
the standard contribution from bubble-wall collisions. In general, the new contribution is suppressed but can become comparable in magnitude when percolation proceeds rapidly or the initial deviation from a spherical geometry is large.

Our mechanism can also be relevant for cosmic string networks, which can arise through spontaneous symmetry breaking in the early Universe~\cite{Kibble:1976sj, Kibble:1980mv, Vilenkin:1981kz}. As the network evolves, strings intersect and intercommute, forming closed loops that oscillate and, as they do so, emit GWs. 
For stable strings, loops are produced continuously from the time of formation until today and the resulting stochastic GW background spans a wide range of frequencies~\cite{Vilenkin:1981bx, Vachaspati:1984gt, Hogan:1984is, Accetta:1988bg}.~\footnote{For comprehensive reviews of cosmic string networks and their cosmological and GW phenomenology, see e.g. \cite{Albrecht:1989mk,Hindmarsh:1994re,Sakellariadou:2009ev, Copeland:2011dx,Schmitz:2024gds}.} Currently the strongest bounds on such backgrounds from local cosmic strings are provided by pulsar timing arrays (PTA). The latest NANOGrav 15‑year dataset~\cite{NANOGrav:2023hvm} disfavours standard, stable cosmic strings as the source of the GW background recently reported by PTAs~\cite{NANOGrav:2023gor, EPTA:2023fyk, Reardon:2023gzh, Xu:2023wog}, and places a stringent upper limit $G\, \mu_s\lesssim10^{-10}$ on the string tension $\mu_s$ (measured in units of Newton’s constant $G$). This bound is significantly stronger than those derived from cosmic-microwave-background (CMB) observations, which are at the level of~\cite{Charnock:2016nzm} $G \mu_s\lesssim10^{-7}$. 

However, if the strings are metastable and persist only up to some intermediate epoch, the low‐frequency part of the GW spectrum is suppressed and the above-mentioned PTA (and CMB) bounds can be evaded. 

As mentioned above, one well-known decay mechanism of local cosmic strings involves the nucleation of monopole–antimonopole pairs along the string core~\cite{Vilenkin:1982hm}, a process whose GW signatures were discussed for example in~\cite{Martin:1996ea,Martin:1996cp,Leblond:2009fq,Gouttenoire:2019kij,Buchmuller:2021mbb, Servant:2023tua} (see~\cite{NANOGrav:2023hvm, Fu:2023mdu, Buchmuller:2023aus} for an analysis in the context of the latest NANOGrav data).
In contrast, our scenario introduces a qualitatively new decay channel: instead of fragmenting through monopole pair production, the string network decays via $O(2) \times O(2)$ bubble nucleation. This process leads to unique features in the GW spectrum, making it potentially distinguishable from the monopole-induced scenario. 
While a complete comparative study of the GW signatures in different models is left for future work, here, we explore the possibility that the GW background observed by PTAs could arise from a string network that decays in a first-order phase transition. To that end, we fit the GW spectrum predicted by our model to the NANOGrav 15-year data.  

Finally, we discuss a more comprehensive cosmological framework that explicitly includes the formation dynamics of the string network. The model features two interacting scalar fields, where a real scalar triggers the false vacuum decay of a complex scalar some time after the strings have formed.

The paper is organized as follows. In Sec.~\ref{sec:minimal_setup}, we define our minimal setup based on a complex scalar field with a sextic potential, charged under either a local or global $U(1)$ symmetry. Sec.~\ref{sec:thin-wall} presents a detailed review of the thin-wall limit for both the cosmic string and the standard Coleman bounce. In particular, we provide a derivation of the bounce action by treating the bubble wall radius as the dynamical variable. In Sec.~\ref{sec:string_induced}, we apply this formalism to the string-induced bounce, computed both for a global and local $U(1)$ symmetry, providing us with an explicit expression for the bounce action that determines the tunneling rate.  Section~\ref{sec:pheno} discusses the phenomenological implications of our results, including a comparison between the tunneling rates in the string-induced and string-less cases in cosmological space-times, a derivation of the quadrupole moment of the $O(2) \times O(2)$ bubble, and a first estimate of the GW spectrum produced by a cosmic string network that features this type of metastability. We also outline a more complete cosmological scenario that includes the formation of the string network. We conclude in Sec.~\ref{sec:conclusion}.

\section{Minimal set-up} \label{sec:minimal_setup}

For our minimal setup, we consider a complex scalar field charged under either global or local $U(1)$ symmetry. We further require the existence of a symmetry-breaking false minimum and a symmetric true minimum separated by a potential barrier. 
This setup due to its nontrivial vacuum manifold allows for the existence of cosmic strings. At the same time, the metastability of the false vacuum renders the cosmic string  unstable against quantum tunneling, which proceeds via the nucleation of bubbles of the true vacuum. This vacuum decay, as we will discuss, can either be dominated by conventional $O(4)$ bubbles \`a la Coleman or by a new class of $O(2) \times O(2)$ symmetric bubbles, which are aligned with the cosmic string.

The Lagrangian density can be written as\footnote{We work in units with $\hbar=c=1$ and adopt the mostly-plus metric signature.}
\beq \label{eq:Lagrangian}
\mathcal{L}(\phi, A) = - [D_{\mu} \phi ]^{*}[D_{\mu} \phi ]  - V(\phi)  - \frac{1}{4}  {F_{\mu\nu}F^{\mu \nu}}\,,
\eeq
where $F_{\mu\nu} = \partial_{\mu} A_{\nu} - \partial_{\nu} A_{\mu}$ is the field strength associated with the gauge potential $A_\mu$, and $D_{\mu} = \partial_{\mu} - i g A_{\mu}$
denotes the covariant derivative with gauge coupling $g$.

\begin{figure}[tbp]
    \centering
    \includegraphics[width=0.45\textwidth]{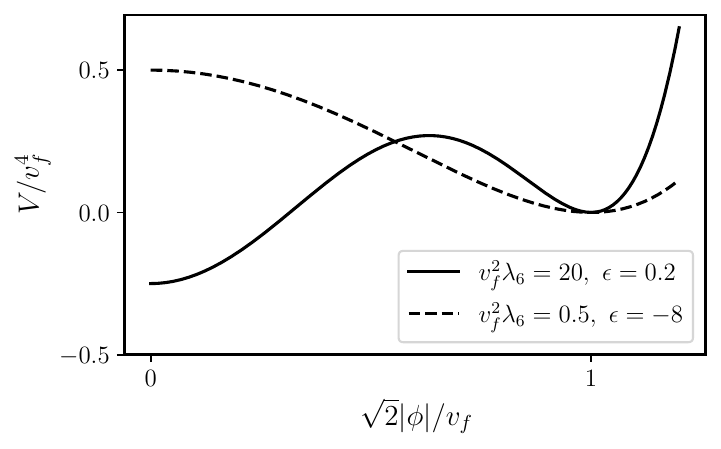}
    \caption{Sketch of the potential in Eq.~\eqref{eq:V2}. For $\epsilon<0$, the true vacuum lies at $|\phi| = v_f/\sqrt{2}$ and the system admits stable cosmic strings (dashed line).  In this work, we focus on the case $\epsilon > 0$, where the true vacuum is at $\phi = 0$, rendering the cosmic string metastable (solid line).}
    \label{fig:V}
\end{figure}

As an explicit example, we consider a potential
\begin{equation}
\label{eq:V}
V(\phi) =  V_1 + \mu^2 \phi^{*} \phi - \lambda (\phi^{*} \phi )^2 + \lambda_6(\phi^{*} \phi )^3\,,
\end{equation}
where $V_1$ is a field-independent constant, $\mu^2 > 0$ a mass parameter, $\lambda >0$ a dimensionless quartic coupling, and $\lambda_6 > 0$ the coefficient of the (irrelevant) sextic operator. The latter has mass dimension $-2$ and arises naturally in an effective field theory (EFT) framework. For simplicity, we omit higher-dimensional operators, although they could be included with only mild restrictions on their coefficients.
The condition 
\begin{align}
\label{eq:bounds}
1/4<\frac{\lambda_6 \mu^2}{\lambda^2}<1/3\,,
\end{align}

which is compatible with order unity choices for $\lambda$ and $\lambda_6 \mu^2$, then ensures the existence of a symmetry-breaking false minimum at $|\phi|=v_f/\sqrt{2}$, where 
\begin{align}\label{eq:vf}
v_f^2=\frac{2}{3 \lambda_6}\left(\lambda + \sqrt{\lambda^2 -  3 \lambda_6 \mu^2}\right)\,.
\end{align}
It is  separated from the symmetric true minimum at $\phi=0$ by a potential barrier (see Fig.~\ref{fig:V} for an explicit example). 
Decomposing the field into a real amplitude $f(x)/\sqrt{2}$ and complex phase $\varphi(x)$, 
\beq
\phi(x) = \frac{f(x)}{\sqrt{2}}e^{i\varphi(x)}\,,
\eeq
and choosing $V_1$ such that the false minimum has vanishing potential energy, we obtain
\beq\label{eq:V2}
V(f) = \frac{\lambda_6}{8}(f^2 -\epsilon v_f^2 )(f^2-v_f^2)^2 \,,
\eeq
where we introduced the dimensionless parameter
\begin{align}\label{eq:epsilon}
\epsilon = -2 + \frac{\lambda \left(4 \lambda - 3 \lambda_6 v_f^2 \right)}{2 \lambda_6 \mu^2}\,,
\end{align}
which controls the energy difference  $\Delta V$ between the false and the true minimum,
\beq
    \Delta V = V(v_f) - V(0) =\frac{\epsilon \, \lambda_6 \, v_f^6}{8} \,.
\eeq
Demanding that $v_f \in \mathbb{R}$ and $\epsilon > 0$, then recovers \eqref{eq:bounds}.
We will later see that many aspects of our discussion are rather universal as they will only depend on the effective parameters $\Delta V$, $\epsilon$, and the \textit{wall tension} $\sigma$. The latter is given by the integral
    \beq\label{eq:tension}
    \sigma=\int_0^{v_f} df \sqrt{2 V(f)}\,,
    \eeq 
and evaluates to $\sigma = (\sqrt{\lambda_6}/8)v_f^4$ for $\epsilon \to 0$ in the case of our explicit example.

\section{Thin-wall limit of strings and bounces} 
\label{sec:thin-wall}

We will first discuss the thin-wall limit of (straight) cosmic strings and bounce solutions, which both are extended field configurations solving the full nonlinear equations of motion with $ O(2)$ and $O(4)$ symmetry, respectively. While the thin-wall approximation is widely used in the literature to derive bounce solutions, it has been less frequently applied to cosmic strings. We therefore examine its regime of validity in detail for both the global and local cases. Geometrically, this limit corresponds to approximating the string core by a cylinder, which is hollow in the global case, and filled with a magnetic field aligned with the cylinder axis in the local case. This analysis sets the stage for discussing a new class of instanton solutions with $O(2) \times O(2)$ symmetry, which can be understood as generalizing \textit{both} the cosmic string and bounce solutions. 
The reader familiar with these topics can skip directly to Sec.~\ref{sec:string_induced}.

\subsection{Cosmic string} 
Cosmic strings are topological defects, similar to domain walls or monopoles, that form when the vacuum manifold $\mathcal{M}$ admits non-contractible loops.\footnote{More formally, this requires a non-trivial first homotopy group of the (true or false) vacuum manifold, $\pi_1(\mathcal{M}) \neq 0$.} In field theory, a canonical realization is provided by a complex scalar field with a spontaneously broken $U(1)$ symmetry~\cite{Nielsen:1973cs}. Away from the string core, the field acquires a nonzero expectation value, $|\phi|\simeq v_f/\sqrt{2}$. In terms of the scalar field potential introduced above, this corresponds to either the global (dashed) or local (solid) minimum in Fig.~\ref{fig:V}.
As mentioned before, in this work we are mostly interested in the latter, metastable case.  As its defining property, the cosmic string has a non-vanishing winding number $n$ associated with its complex phase. Correspondingly, the (straight) cosmic string can be expressed as $\phi \simeq (v_f/\sqrt{2}) e^{in\theta}$ sufficiently far away from the string core. Here, we adopted cylindrical coordinates $x^{\mu} = (t,\rs,\theta,z)$, where $\theta$ denotes the polar angle and $\rs = \sqrt{x^2 + y^2}$ the radial distance from the string core at $r=0$.  The $z$-axis is aligned with the string. Cosmic strings are topological solitons which cannot be continuously ``un-winded''. In this work, we will assume that the string tension $\mu_s$ is small such that we can neglect its gravitational back-reaction. In particular, taking into account the string's deficit angle of order $8 \pi\, G \,\mu_s \ll 1$ would not alter the results of this work significantly. 

In its ground state the string profile has a cylindrical $O(2)$ symmetry. We make the static ansatz

\begin{align}\label{ansatz_a}
A_\theta(x) =   \frac{n}{g}\, a(r) \quad \text{and} \quad A_r=A_t=A_z=0  \,.
\end{align}

We further decompose
\beq\label{eq:ansatz_cosmis_string}
\phi(x) = \frac{f(\rs)}{\sqrt{2}}e^{i \theta n}\,, 
\eeq
where the function $f(\rs)$ denotes the radial string profile.
Inserting the ansatz into \eqref{eq:Lagrangian}, we obtain
\begin{multline}\label{eq:action}
\frac{S_E[f,a]}{TL} = 2 \pi \, \int d\rs \, \rs \Bigg[ \frac{1}{2} f'^2  +V(f) + \frac{1}{2} \, \frac{n^2}{g^2} \,\frac{a'^2}{\rs^2} \\
+ \frac{1}{2}(1 -a)^2\,n^2\,\frac{f^2}{\rs^2} \Bigg]\,,
\end{multline}
where the Euclidean action is defined as $S_{E}=-iS$.

The equations of motion for $f$ and $a$ then become
\begin{subequations} \label{eq:eom_cosmic_string}
\begin{align}
\label{string_eq}
f'' + \frac{1}{\rs}f' - \frac{n^2}{\rs^2}(1 - a)^2f - \frac{dV}{df} =0\,.\\
a''-\frac{1}{r}a' + (1-a)\, g^2 f^2 =0\,.
\end{align}
They have to be supplemented with the boundary conditions 
\begin{align}
\label{boundary_string}
f(\rs \rightarrow \infty) &= v_f\,, &\quad f'(0) &= 0\,, \nonumber \\
a(\rs \rightarrow \infty) &= 1 \,, & a'(0) &= 0\,.
\end{align}
\end{subequations}
In other words, the string configuration interpolates between the symmetric and the broken phase in the interior and exterior of the string, respectively. We note that there is a cosmic string configuration irrespective of whether there is a maximum or minimum at $f=0$ (respective cases corresponding to the dashed and solid line in Fig.~\ref{fig:V}).
While the above system describes a local cosmic string, we can readily recover the global case by taking $g \to 0$ and $a/g \to 0 $. 

In the \textit{global} case, we infer from the second row in \eqref{eq:action} that the energy per length of the string diverges logarithmically with the radial distance in the exterior; explicitly, we find  
\begin{align}
\left(\frac{E}{L}\right) 
\supset {\pi} \int_{R_\mathrm{min}}^{R_\mathrm{max}} d \rs \, n^2\,\frac{f^2}{r} \nonumber 
\sim \pi\,n^2 v_f ^2 \,\ln \left( \frac{R_\mathrm{max}}{R_\mathrm{min}} \right)\,,
\end{align}
where we introduced the infrared cutoff scale $R_{\mathrm{max}}$, and $R_\mathrm{min}$ is chosen to be sufficiently far outside the string core such that $f \simeq v_f$. In realistic scenarios, $R_{\mathrm{max}}$ can be identified with the typical string separation.

For \textit{local} strings, the screening by the gauge fields prevents this divergence. Instead, a quantized magnetic flux, aligned with the string, is generated in the interior.  The gauge field takes the form such that $D_\mu\, \phi \simeq 0$ outside the string core, implying
\beq\label{eq:local_string_condition}
A_{\mu} \simeq \frac{1}{ig} \partial_{\mu} \ln \phi \,.
\eeq
As a result, the winding around the string translates into a non-vanishing magnetic flux along the string quantized in units of $2 \pi / g$. Indeed, we can re-write $n$ through
\beq
\begin{split}
    n = \frac{1}{2\pi} \int_{0}^{2\pi} \frac{\partial \ln \phi }{\partial \theta} d\theta = \frac{g}{2\pi} \oint \mathbf{A} \,d\mathbf{l} = \frac{g}{2\pi} \int \mathbf{B} \,d\mathbf{S} \,,
\end{split} 
\eeq
where the magnetic field is defined as $B^i =- 1/2 \,\epsilon^{ijk}\, F_{jk}$ with (i,\,j,\,k = 1,2,3), and we used Stokes' theorem for the last equality. Using our cylindrically symmetric ansatz, the condition \eqref{eq:local_string_condition} translates to $a\simeq -(i/n)\,\partial_{\theta} \ln \phi=1$ in accordance with the boundary condition \eqref{boundary_string}. From the second line in \eqref{eq:action}, we see that in this case the winding and gauge field contributions indeed cancel.

We now apply the thin-wall limit, which is applicable when the energy difference between the two minima $\Delta V$ is much smaller than the height of the potential barrier $V_\mathrm{max}$. While the discussion before only demanded the existence of a local minimum at $v_f$, the presence of a potential barrier now also requires the symmetric phase at $f=0$ to be a minimum (corresponding to the solid line in Fig.~\ref{fig:V}).   In this limit, the field profile $f(r)$ undergoes a sharp transition from $f=0$ to $f=v_f$ within the `wall' region $ \Rs-\Delta R < \rs <  \Rs + \Delta R$ with thickness $\Delta R \ll \Rs$.\footnote{Qualitatively, this can be understood by formally replacing $r \to t$ in \eqref{string_eq} and considering the dynamical problem in the `up-side-down' potential with time-dependent damping and mass term.} 
Multiplying \eqref{string_eq} by $f'$ and integrating it over $\rs$ then leads to 
\begin{multline}
\label{eq:f_prime_thin_wall}
f'^2 - 2V(f) \simeq \frac{1}{\Rs}\int_0^f d\tilde f \,\tilde f' +  \frac{n^2}{2\Rs^2} \left[ 1-a(\Rs)\right]^2f^2\,,
\end{multline}
where we approximated $f'=0$ for $|\rs-\Rs| \gg \Delta R $, used $V(f(0)) \simeq 0 $, and dropped terms of order $\Delta R/\Rs \ll 1$ in accordance with the thin-wall approximation.
In addition, we also neglect the terms on the right-hand side of \eqref{eq:f_prime_thin_wall}, which implies 
\beq
\label{eq:wall_eq}
f'^2 \simeq 2V(f)\,.
\eeq
For $\Delta V = 0 $, this equation determines the wall profile function $\fw(\chi)$ with the boundary condition $V(\fw(0))=V_\mathrm{max}$ such that $f(\rs)\simeq f_\mathrm{wall}(r-\Rs)$. To be specific, for our reference potential in \eqref{eq:V2}, we derive (for $\epsilon=0$)
\begin{align}\label{eq:wall_profile}
f_\mathrm{wall}(\chi)= \frac{v_f\, e^{\frac{1}{2} \sqrt{\lambda_6 } \, \chi \,v_f^2}}{\sqrt{e^{\sqrt{\lambda_6 }\, \chi \,v_f^2}+2}}\,,
\end{align}
which indeed interpolates between $f=0$ and $f=v_f$ within an interval set by the scale $\Delta R =  1/\sqrt{v_f^4\,\lambda_6 }$.

We will now check \textit{a posteriori} that the approximations leading to \eqref{eq:wall_eq} constitute a self-consistent choice within the thin-wall limit. To that end, we evaluate each of the two terms on the right-hand side of \eqref{eq:f_prime_thin_wall} at $f(\Rs)$  and demand that they are small compared to $V_\mathrm{max} \equiv V(f(\Rs))$.  For the first term, we obtain
\begin{align}
\frac{1}{\Rs}\int_0^{f(\Rs)} d\tilde f \,\tilde f'\simeq \frac{1}{\Rs}\int_0^{f(\Rs)} d \tilde f \sqrt{2V} \simeq \frac{\sigma}{2\Rs}\,,
\end{align}
where we  used the definition in \eqref{eq:tension} together with \eqref{eq:wall_eq}. Similarly, the second term evaluates to
\begin{align}
\frac{n^2}{2\Rs^2} \left[ 1-a(R)\right]^2f(\Rs)^2\lesssim \frac{n^2 f(\Rs)^2}{2 \,\Rs^2}\,.
\end{align}

The validity of the thin-wall approximation thus translates into the two constraints
\begin{align}\label{eq:thin_wall_conditions}
 \left\{\frac{n^2 \,\left[ 1-a(R)\right]^2f(\Rs)^2}{2 \,\Rs^2},\,\frac{\sigma}{2\Rs} \right\}\ll V_\mathrm{max}\,.
\end{align}

Let us first consider the global cosmic string ($a/g=g=0$). We evaluate the Euclidean action in \eqref{eq:action} in the thin-wall  limit, which provides us with an expression for the energy per length $E(\Rs)/L = S_E(\Rs)/(TL)$ of the cosmic string,
\begin{equation}
\label{S_thinwall_string}
 \frac{E(\Rs)}{L} =
- \underbrace{\pi \Rs^2 \Delta V}_{\text{interior}} 
+ \underbrace{2 \pi \Rs \sigma}_{\text{wall}} 
+ \underbrace{  \pi\, n^2 v_f ^2 \,\ln \Bigl( \frac{R_{\rm max}}{\Rs} \Bigr)   }_{\text{exterior}}\,.
\end{equation}
Here, we indicated how different terms arise from the radial integration, which we split into an `interior' ($r<R-\Delta R$), `wall' ($R-\Delta R<r<R+\Delta R$), and `exterior' ($R+\Delta R<r$) region.
The energy function has a local minimum at
\beq
\label{eq:metastable_string}
R_s = \frac{\sigma}{2\Delta V} (1 - \sqrt{1 - x})\,,
\eeq
where the dimensionless parameter $x$ is defined as
\beq\label{def:x}
x = \frac{2n^2 v_f^2 \Delta V}{\sigma^2}\,.
\eeq
Such a local minimum exists only if $0<x<1$ and describes a classically stable string. For $x>1$ the string is classically unstable against expanding its radius, which is referred to as dissociation.

The energy per length of the metastable string is the string tension, $\mu_s = E(R_s)/L$. For $R_{\rm max} \gg R_s$ it is dominated by the exterior contribution and given by $\mu_s \simeq  \pi n^2 v_f^2 \ln ( {R_{\rm max}}/{R_s} )$ for all values of $x$.

\begin{figure}
    \centering
    \includegraphics[width=0.99\linewidth]{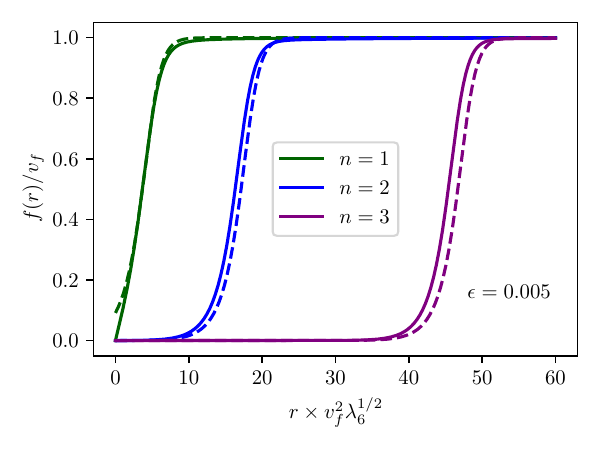}
    \caption{Field profiles of global strings for the potential in~(\ref{eq:V2}) with $\epsilon =0.005$. We compare the full numerical solution, obtained using the gradient flow method (solid lines), with the thin-wall approximation (dashed lines) and find good agreement.}
    \label{fig:profiles}
\end{figure}

Having an explicit expression for the string radius $R_s$ allows us to derive parameter constraints from the thin-wall approximation. To keep the discussion concrete, we consider the potential of our working example in \eqref{eq:V2}. Substituting $\sigma = (\sqrt{\lambda_6}/8)v_f^4$, $V_\mathrm{max}=v_f^6\,\lambda_6/54$, and $\Delta V = v_f^6 \, \lambda_6 \, \epsilon /8$ into   \eqref{eq:metastable_string} and \eqref{eq:thin_wall_conditions}, we derive the two conditions
\begin{align}
\epsilon \left(1-\sqrt{1-x} \right)^{-1} &\ll \frac{8}{27}\,,\\
n^2\,\epsilon^2 \left(1-\sqrt{1-x} \right) &\ll \frac{1}{36}\,,
\end{align}
where $x=16\, n^2 \epsilon$.
The first inequality also guarantees $\Delta R/R \ll 1$, as assumed above. It is useful to distinguish two regimes. For $x\lesssim 1$, i.e.\ $\epsilon \lesssim 1/(16 n^2)$, the conditions reduce to $n^2 \gg 0.21$. For $x\ll 1$, we instead require $\epsilon \ll 1/(16 n^2)$ and $n^2 \gg 0.42$. Thus, our analysis is quantitatively reliable only for winding numbers $n \geq 3$, although we expect the qualitative results to extend to the unit winding case. This would suggest our analysis is only valid for large enough winding number; however, we show in Fig.~\ref{fig:profiles} that even for unit winding the thin-wall profile (\ref{eq:wall_profile}) (dashed) provides a good description of the full numerical solution (solid). Here, the numerical solutions are obtained via the gradient flow method, where we promote $f(\rs) \rightarrow f(\rs, s) $ and `evolve' the profile according to
\beq
\frac{\partial f }{\partial s} = f'' + \frac{1}{\rs}f' - \frac{n^2}{\rs^2}f - \frac{dV}{df} ,
\eeq
subject to the appropriate boundary conditions and starting from some guessed initial configuration. As the evolution parameter $s$ increases, the configuration relaxes toward a `static' solution, which by construction corresponds to the correct string profile. 

We note that the above bounds on the winding $n$ can be rewritten as  $\epsilon \ll 1 $ (or equivalently as $\Delta V/V_\mathrm{max} \ll 1$), which is precisely the standard thin-wall requirement. Moreover, the precise bounds should be expected to change for potentials with different shapes.

For the local cosmic string, we again assume that the scalar field profile is given by the solution of \eqref{eq:wall_eq}, e.g.\ $f(\rs)\simeq f_\mathrm{wall}(r-\Rs)$, and in addition set
\begin{align}\label{eq:ansatz_a_thin_wall}
a(r) = 
\begin{cases}
\frac{r^2}{R^2} & \text{for} \quad r<R-\Delta R\,,\\
1  & \text{for} \quad  r>R+\Delta R\,.
\end{cases}
\end{align}
where $a(r)$ is assumed to smoothly interpolate between both regime in the wall region. With these choices it is straightforward to check that the equations of motion in \eqref{eq:eom_cosmic_string} are solved both in the interior and exterior (if we use that $f\simeq 0$ for $r<R-\Delta R$). The energy functional then evaluates to
\begin{equation}
\label{S_thinwall_string_local}
 \frac{E(\Rs)}{L} =
- \underbrace{\pi \Rs^2 \Delta V\, + \frac{2\pi \,n^2}{g^2}\frac{1}{R^2}}_{\text{interior}} 
+ \underbrace{2 \pi \Rs \,\sigma}_{\text{wall}} \,,
\end{equation}
where in contrast to the global string there is no exterior contribution as the winding and flux terms exactly cancel as anticipated. Introducing the dimensionless parameter
\begin{align}\label{def:y}
y=\left(\frac{8}{3}\right)^3\frac{n^2\Delta V^3 }{g^2 \sigma^4}\,,
\end{align}
we find that a minimum of $E(R)$ exists for $0<y<1$ and takes the form 
\begin{align}
R_s = \frac{3\, \sigma}{4\,\Delta V} h(y)\,,
\end{align}
where $h(y)$ interpolates monotonically between $h(0)=0$ and $h(1)=1$.\footnote{For completeness, it takes the form 
{\small
\begin{align}
h(y)=\frac{1}{6} \left(\sqrt{6 z+4}-\sqrt{-6 z+8+8 \sqrt{2/(3z+2)}}+2\right)\,, \nonumber
\end{align}
}
where $z(y)$ is determined by solving $y=z^3/(2+3z)$.
} The string tension is of the order $\mu_s \sim (\sigma/g)^{2/3}$ for all values of $y$. For small $y$ the term proportional to $\Delta V$ in \eqref{S_thinwall_string_local} provides only a subdominant contribution to $\mu_s$.

The thin-wall conditions in \eqref{eq:thin_wall_conditions} for the sextic potential now translate into
\begin{align}\label{eq:bound_h}
\epsilon \,h(y) \ll \frac{8}{9}
\end{align}
where $y=(16\,\epsilon/3)^3 \, n^2 (v_f^2 \lambda_6)/g^2$\,. Taking both the $y \to 1$ and $y \ll 1$ limit of \eqref{eq:bound_h}, we derive the sufficient condition 
\begin{align}
 \left(\frac{g^2}{n^2 v_f^2 \lambda_6}\right)^{1/3} \ll 1
\end{align}
for the thin-wall approximation to be satisfied.

In summary, we have shown that both for the global and local string we can apply the thin-wall approximation. While in the former case, the limit is only marginally satisfied for a unit winding string, in the latter case, it can be more easily achieved for a sufficiently small gauge coupling $g \ll 1$ (assuming $v_f\,\lambda_6\sim 1$).

\subsection{Bounce}

Here, we start by reviewing the standard derivation of the bounce action by Coleman~\cite{Coleman:1977py}. The bounce~$\phi_\mathrm{b}(\rho)$ is a real-valued solution to the classical field equations in Euclidean time $\tau = i\,t$ with $\mathcal{O}(4)$ symmetry. It depends on the generalized radial coordinate $\rho = \sqrt{x^2 + y^2 + z^2 + \tau^2}$ and determines the leading contribution to the false vacuum decay rate $\Gamma\propto e^{-B}$ in the semiclassical limit, where 
\begin{align}
\label{eq:bounce_action_normalized}
B = { S_E[\phi_{\rm b}] - S_E[\phi_\mathrm{f.v.}]}\,,
\end{align}
is the normalized bounce action and
$|\phi_\mathrm{f.v.}|=v_f/\sqrt{2}$ is the vacuum expectation value in the false vacuum. In particular, in the absence of a cosmic string, the vacuum configuration takes the same field value everywhere. Inserting the ansatz
\beq\label{eq:ansatz_bounce}
\phi_\mathrm{b}(\rho) = \frac{f(\rho )}{\sqrt{2}}, \quad A_\mu = 0
\eeq
into \eqref{eq:Lagrangian}, we find 

\begin{align}\label{eq:action_bounce}
S_E[f] = 2 \,\pi^2 \, \int d\rho \, \rho^3 \Bigg[ \frac{1}{2} f'^2  +V(f) \Bigg]\,.
\end{align}
The corresponding field equation reads
\beq
\label{bounce_eq}
f'' + \frac{3}{\rho}f' - \frac{dV}{df} =0,
\eeq
which is supplemented with the boundary condition
\beq 
f(\rho \rightarrow \infty) \rightarrow v_f\,, \quad f'(0) = 0\,.
\eeq
The solution correspond to bubbles of true vacuum surrounded by the false vacuum phase.  
Note the (formal) similarity between the field equations (\ref{bounce_eq}) for the bounce and (\ref{string_eq}) for the string. As a result, also here, we can apply the thin-wall limit 
where $f(\rho)$ undergoes a sharp transition from $f=0$ to $f=v_f$ within the `bubble wall' region $ \mathcal{R}-\Delta \mathcal{R} < \rho <  \mathcal{R} + \Delta \mathcal{R}$ with thickness $\Delta \mathcal{R} \ll \mathcal{R}$. Multiplying \eqref{bounce_eq} by $f'$ and integrating over $\rho$ then yields as before  
\beq
\label{eq:wall_eq_2}
f'^2 \simeq 2V(f)\,,
\eeq
where we neglected a term of order $\sigma/(2\mathcal{R}) \ll V_\mathrm{max}$. In particular, the above equation is solved by the \textit{same} wall profile $f_\mathrm{wall}(\chi)$ as before, taking the form \eqref{eq:wall_profile} for the sextic potential. The only difference is that we now identify $f(\rho)\simeq f_\mathrm{wall}(\rho-\mathcal{R})$, {i.e.}, we evaluate the wall profile function with respect to the Euclidean radius $\rho$. Substituting \eqref{eq:wall_eq_2} into \eqref{eq:action_bounce} and splitting the integral into different regions, we obtain
\beq\label{eq:Eucilidian_O4}
S_E(\mathcal{R}) = - \underbrace{V_{3}(\mathcal{R}) \Delta V}_{\text{interior}} 
+ \underbrace{A_{3}(\mathcal{R}) \sigma }_{\text{wall}} \,,
\eeq
where $V_{3}(\mathcal{R}) = \frac{1}{2} \pi^2 \mathcal{R}^4$ and $A_{3}(\mathcal{R}) = 2 \pi^2 \mathcal{R}^3$ are the volume and the surface area of the $3$-sphere, respectively. The `critical' radius $\mathcal{R}_c=3\sigma/\Delta V$ can then be determined by extremizing the action. Plugging this back into \eqref{eq:Eucilidian_O4} recovers the standard result
\begin{align}\label{eq:Coleman_bounce}
B_0= \frac{27\,\pi^2 }{2}\frac{\sigma^4}{\Delta V^3}\,.
\end{align}
The constant value of the critical radius in four-dimensional Euclidean space $\mathcal{R}_c = \sqrt{R_b(\tau)^2+\tau^2}$ then determines the radial trajectory $R_b(\tau)$ of the bubble wall after nucleation. Solving for $R_b(\tau)$, we find 
\begin{align}\label{eq:bounce_solution}
R_b(\tau) = \sqrt{\mathcal{R}_c^2 -\tau^2}\,,
\end{align}
which can be rotated back to real time, $\tau = i\,t$.
\bigskip

We now demonstrate how the same result can be obtained if one assumes a lower degree of symmetry for the bounce (see for example~\cite{Ai:2020vhx}). To that end, we  assume only a spatial $\mathcal{O}(3)$ symmetry and make a more general ansatz for the field profile in~\eqref{eq:ansatz_bounce} by replacing $f(\rho) \to f(\tau, |\mathbf{x}|)$, where $|\mathbf{x}|=\sqrt{x^2+y^2+z^2}$. The equation of motion then reads
\begin{align}
\label{bounce_eq_O3}
\partial_\tau^2f + \partial_{|\mathbf{x}|}^2 f + \frac{2}{|\mathbf{x}|}\partial_{|\mathbf{x}|}f - \frac{dV}{df} =0\,.
\end{align}
We make the thin-wall ansatz
\beq
\label{ansatz:O3xt}
f(\tau, |\mathbf{x}|) = \fw( \gamma(\tau )(|\mathbf{x}|-R(\tau)) ) \,,
\eeq
where $\fw(\chi)$ solves \eqref{eq:wall_eq_2} and is given by~\eqref{eq:wall_profile} for our sextic potential. Crucially, we introduced the factor 
\begin{align}\label{eq:gamma}
\gamma=\frac{1}{\sqrt{1+\dot R^2}}\,.
\end{align}
where $\dot R = dR(\tau)/d\tau$. Geometrically, multiplying the radial distance by the $\gamma$-factor gives the (proper) distance from the wall surface in the four-dimensional Euclidean spacetime. It is nothing else than the Lorentz factor, which accounts for the fact that as the bubble wall moves with some velocity $\dot R = dR/d\tau$ its width gets contracted compared to a static wall with the same radius, which would be described as $\fw(|\mathbf{x}|-R )$. It is straightforward to show that the ansatz \eqref{ansatz:O3xt} satisfies \eqref{bounce_eq_O3} if we assume that $f_\mathrm{wall}'\simeq f_\mathrm{wall}'' \simeq 0$ outside the wall region and $ \ddot R\, \Delta \mathcal{R}\ll1$. As a consistency check, we will show \textit{a posteriori} that the last relation is indeed satisfied.

With this ansatz the action in \eqref{eq:Lagrangian} takes the form,
\beq
\label{S_thinwall_bounce}
S_E[R(\tau)] = \int d\tau \Bigl[   - \underbrace{ \frac{4\pi}{3}R^3(\tau) \Delta V }_{\text{interior}} 
+ \underbrace{ 4\pi R^2(\tau) \sigma \, \gamma^{-1} }_{\text{wall}} \Bigr]\,.
\eeq 
where we used the thin-wall approximation.
In other words, we have reduced the field theoretic problem in \eqref{eq:Lagrangian} to a quantum mechanical problem described by the dynamical variable $R(\tau)$. Its equations of motion is
\begin{align}
 \ddot{R}+\gamma ^{-3}\frac{3}{\, \mathcal{R}_c}- \gamma^{-2}\frac{2}{ \, R(t)}=0\,,
 \label{bounce:eom_R}
\end{align}
and the boundary conditions read
\beq
\begin{split}
&R(\tau \rightarrow \pm \infty )=0 \: \: \: (\text{the field is the false vacuum})\,,\\
&\frac{dR}{d\tau}(\tau =  0) = 0 \: \: \: \:  \: \: (\text{regularity})\,.
\end{split}
\eeq
It is straightforward to check that this system is solved by the bounce solution in \eqref{eq:bounce_solution}, which was derived by making an  $\mathcal{O}(4)$-symmetric ansatz. Moreover, from \eqref{bounce:eom_R} we see that the assumption $\ddot R\, \Delta \mathcal{R}\ll1$ was indeed satisfied in the thin-wall limit. In the next section, we will make use of a similar approach to make the bounce calculation analytically feasible.

\section{String-induced bubble nucleation}
\label{sec:string_induced}
In this section, we discuss tunneling induced by a single, straight cosmic string. As before, we align the string along the $z$-direction. Describing vacuum decay in the presence of a cosmic string requires finding the bounce solution to the classical field equations. Due to the presence of the cosmic string the symmetry of the bounce configuration is reduced according to the pattern
\begin{align}\label{eq:pattern}
\underbrace{O(4)}_{\rho} \underset{\text{string}}{\quad \longrightarrow \quad}\underbrace{O(2)}_{\varrho} \, \times \, \underbrace{O(2)}_{r}\,,
\end{align}
where the first $O(2)$ corresponds to a radial symmetry in the two-dimensional Euclidean subspace with radius  $\varrho = \sqrt{z^2+\tau^2}$ and the second $O(2)$ is the cylindrical symmetry around the string with polar coordinate $\rs = \sqrt{x^2+y^2}$. As before, we will first discuss the simpler global case in Sec~\ref{sec:global_string} before moving on to the local string in Sec~\ref{sec:local_string}.

\subsection{Global cosmic string}
\label{sec:global_string}
According to \eqref{eq:pattern}, we look for solutions of the form
\beq\label{eq:ansatz_O2_O2}
\phi(x) = \frac{ f( \varrho, r)}{\sqrt{2}}e^{i \theta n}\,,
\eeq
which reduces to the ansatz in \eqref{eq:ansatz_cosmis_string} when the $\varrho$ dependence is dropped. With this, the Euclidean action evaluates to
\begin{multline}
\label{S_E_O2O2_field}
S_E = 4\pi^2 \int d\varrho\, dr  \varrho \,r\,\Bigg[ \frac{1}{2} (\partial_{\varrho} f)^2+\frac{1}{2}(\partial_{r} f)^2 + V(f) \\
+\frac{1}{2} \,n^2\frac{ f^2}{r^2}\Bigg]\,.
\end{multline}
As before, we take the thin-wall limit where the bounce solution can be expressed in terms of the wall profile function $\fw$ defined in \eqref{eq:wall_eq}. We center it around the radial position of the brane at $ r=R(\varrho)$, where the $\varrho$ dependence preserves the O(2) symmetry. Generalizing the ansatz in \eqref{ansatz:O3xt}, we write
\beq
\label{ansatz_O2O2}
f( \varrho, r) = \fw\Bigl( \gamma(\varrho )[r-R(\varrho)] \Bigr),
\eeq
where the definition of the $\gamma$-factor in \eqref{eq:gamma} generalizes to $\gamma^{-1}=\sqrt{1+\dot R^2}$ with $\dot R = dR(\varrho)/d\varrho$.  We stress that a similar approach was employed in~\cite{Lee:2013zca}. However, there the $\gamma$-factor was not included in the ansatz, making their results valid only in the $\dot R\ll 1$ regime for which $\gamma \simeq 1$. We will refer to this later as the `non-relativistic approximation'. Since the expansion of bubble walls reaches relativistic velocities, inclusion of the $\gamma$ factor is crucial for describing the post-nucleation dynamics of the bubble correctly. 

\begin{figure*}[t]
    \centering
    \includegraphics[width=0.47\textwidth]{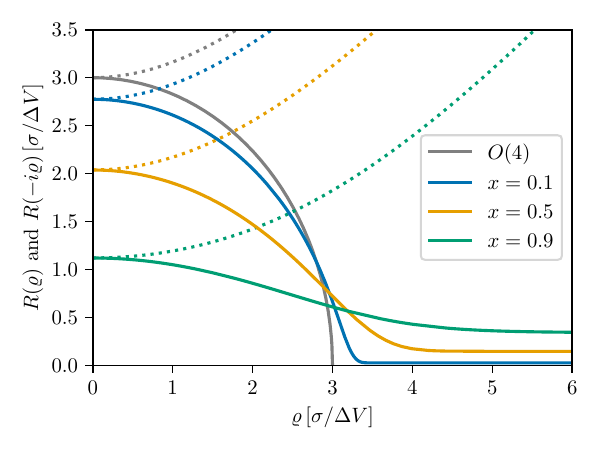}
    \includegraphics[width=0.47\textwidth]{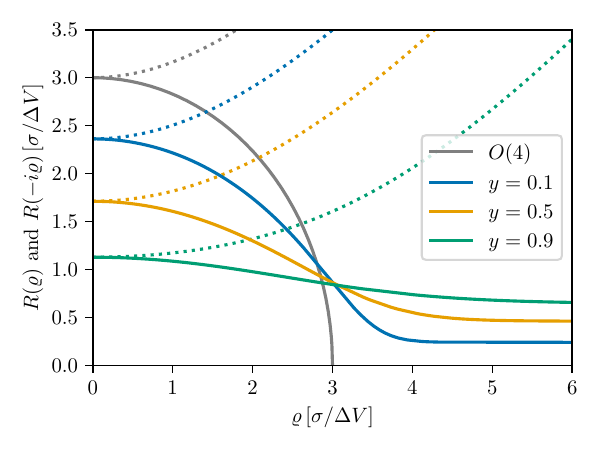}
    \caption{Radial profiles $R(\varrho)$ of the bounce solution for global (left) and local (right) strings with three values of the parameter $x$ or $y$ (solid lines). For $x,y \ll 1$, the profile converges to the $O(4)$-symmetric solution $ \sqrt{\mathcal{R}_c^2 - \varrho^2}$, with $\mathcal{R}_c = 3\sigma/\Delta V$, which represents the Coleman bounce in the absence of strings and is shown as the gray curve. The dashed curves depicts the analytic continuation of each profile to $\varrho\to -\,i\varrho$, describing the late post-phase transition evolution of the bubble wall. All quantities are plotted in units of $\sigma/\Delta V$.}
    \label{fig:bounce_profiles}
\end{figure*}

In any event, substituting \eqref{ansatz_O2O2} into \eqref{S_E_O2O2_field} and applying the thin-wall limit yields
\begin{multline}
\label{S_thinwall_stringbounce}
S_E = 2 \pi \int \varrho \,d\varrho \Bigl[
- \underbrace{\pi R^2(\varrho)\Delta V}_{\text{interior}} \\
+ \underbrace{2\pi \,R(\varrho)\sigma \,\gamma^{-1}(\varrho)}_{\text{wall}} 
+ \underbrace{  \pi \,n^2 v_f^2 \ln \Bigl( \frac{R_{\rm max}}{R(\varrho)} \Bigr)   }_{\text{exterior}}
\Bigr]\,,
\end{multline}
where as before we highlighted the origin of the different terms when performing the radial integration. In addition, we used $\fw'=0$ outside the wall region, i.e.\ for $|r-\Rs| \gg \Delta R $, and dropped terms of order $\Delta R/\Rs \ll 1$.
It is convenient to perform a rescaling
\beq\label{eq:rescale}
R \rightarrow \tilde R = R /\alpha,\quad\varrho \rightarrow \tilde \varrho = \varrho/\alpha,
\eeq
with $\alpha = \sigma/\Delta V$. The action then takes the form
\beq
S_E =  \frac{4 \pi^2 \sigma^4}{\Delta V^3 } \int d \tilde \varrho \, \tilde \varrho \left[ - \frac{1}{2}\tilde R^2  + \frac{\tilde R}{\gamma} + \frac{x}{4}\ln \Bigr(\frac{\tilde R_{\rm max}}{\tilde R} \Bigl)  \right].
\eeq

Apart from an overall factor, the action and hence  the properties of the bounce depend on the potential only through the variable $x$. The corresponding equations of motion, obtained by varying the action, take the form
\beq
 \ddot{\tilde  R} + \gamma^{-3}\Bigl(1+\frac{x}{4\tilde R^2}\Bigr) + \gamma^{-2}\Bigl( \frac{ \dot {\tilde  R}}{\tilde \varrho} - \frac{1}{\tilde R}  \Bigr) = 0,
 \label{bounce:eq}
\eeq
and should be solved in combination with the boundary conditions
\begin{subequations}
\label{boundary_cond_string}
\begin{align}
\tilde R(\varrho \rightarrow  \infty )&=\tilde R_{s} && (\text{metastable string})\\
\dot {\tilde R}(\varrho =  0) &= 0&&  (\text{continuity})
\end{align}
\end{subequations}
As a quick sanity check of the above system, we see that it is solved by the static string with radius $\tilde R_s = (1-\sqrt{1-x})/2$ [see Eq.~\eqref{eq:metastable_string}]. Moreover, for $x=0$ it is solved by $ R(\varrho)=\sqrt{\mathcal{R}_c^2 - \varrho^2}$, with $\mathcal{R}_{c} = 3\sigma/\Delta V$. Using that $R(\varrho)^2 +z^2 = \mathcal{R}_c^2 - \tau^2$, we recover the $O(4)$ symmetric bounce solution in \eqref{eq:bounce_solution}. In other words, for an infinitely thin cosmic string, bubble nucleation proceeds in the usual way through the nucleation of spherical bubbles. This is also visualized in Fig.~\ref{fig:bounce_profiles} (left panel) where we depict numerical solutions for different values of $x$ and the Coleman bounce is depicted as the gray curve. The latter is indeed approached for $x\to 0$, while the bounce profile is more stretched along the string axis in the opposite limit.

These numerical solutions are obtained by solving Eq.~(\ref{bounce:eq}) using a standard shooting method to determine the value $\tilde R(0)$ that recovers the boundary conditions in (\ref{boundary_cond_string}). To that end, we discretize $\varrho$ over the interval $[\varrho_{\rm min}, \varrho_{\rm max}]$. As  $\varrho^{-1}$ diverges in one of the coefficients in (\ref{bounce:eq}) as $\varrho \rightarrow 0$, we start the integration at a small but finite $\varrho_{\rm min}$. For small $\varrho$ the solution admits a series expansion, 
\beq\label{small-rho}
\tilde R(\varrho) = \tilde R_{0} + \tilde{R}_2 \frac{\varrho^2}{2} + \mathcal{O}(\varrho^3),
\eeq
where $\tilde{R}_2 = -(1+x/(4\tilde R_{0}^2) - 1/\tilde R_{0})/2$. We use this analytic form at $\varrho_{\rm min}$ to set our initial conditions. For the shooting criterion we require that $\tilde R(\varrho_{\rm max}) =  \widetilde R_{s}$.

\begin{figure*}[t]
    \centering
    \includegraphics[width=0.47\textwidth]{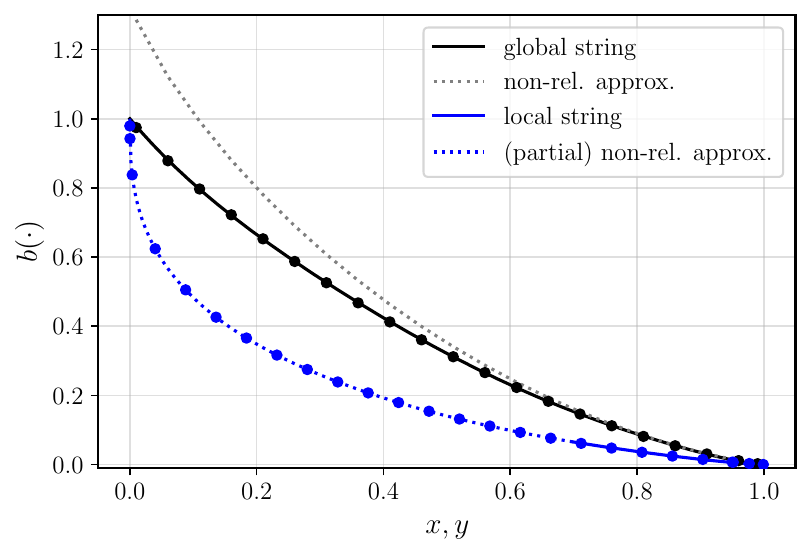}
    \includegraphics[width=0.47\textwidth]{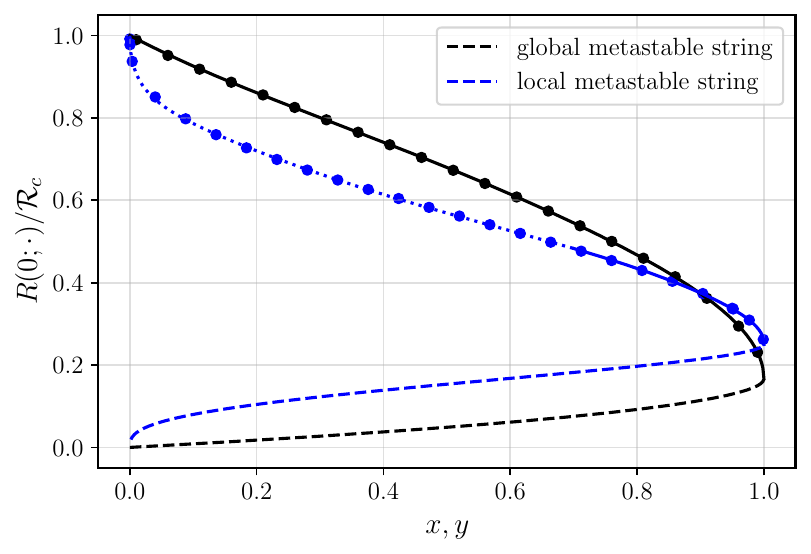}
    \caption{Numerical bounce results for the global (black) and local (blue) string obtained from a shooting algorithm as functions of $x$ and $y$, respectively. The dots represent the numerical shooting result and the solid lines correspond to the semi-analytic fitting functions in \eqref{eq:semi-analytic} and \eqref{eq:semi-analytic_local}. The quantities in both plots are normalized with respect to the Coleman bounce and thus approach unity as $x\rightarrow 0$. 
    {\bf Left:} The normalized bounce action 
    $b(\cdot)=B(\cdot)/B_0$ (solid). The dotted lines depict the `non-relativistic approximation', which fails to recover the Coleman limit for the global string but is reliable for $x \to 1$. 
    {\bf Right:} The  bounce radius evaluated at the bubble center at initial time
    $R(\varrho=0; \cdot)/\mathcal{R}_c$ (solid) and the metastable string radius $R_s(\cdot)/\mathcal{R}_c$ as defined in \eqref{eq:metastable_string} (dashed). For $x,y \to 1$ both quantities approach each other.
    }
    \label{fig:bounce_x_dependence}
\end{figure*}

In the left panel of Fig.~\ref{fig:bounce_x_dependence}, we plot the normalized bounce radius at the center, $R(\varrho =0; x)/\mathcal{R}_c$, as a function of the parameter $x$. The right panel shows the normalized bounce action $b(x) = B(x)/B_0$, where $B(x)$ is the bounce action \eqref{eq:bounce_action_normalized} with the false vacuum contribution (metastable string) subtracted, and $B_0$, defined in \eqref{eq:Coleman_bounce}, is the corresponding  Coleman bounce action in the absence of a string. 
Due to this choice of normalization both quantities approach unity  as $ x\rightarrow 0$ corresponding to the $O(4)$-limit as expected. We stress again that this is not the case for the non-relativistic approach (gray dotted) that assumes $\gamma = 1$. While it still reproduces the right order of magnitude for $b(x)$, the precise value is off by around $25 \%$ when approaching the Coleman bubble limit. On the other hand, for $x\rightarrow 1$, the energy minimum of the metastable string from (\ref{S_thinwall_string}) becomes very shallow and, as a result, the bounce radius only slightly exceeds the string radius and thus $b(x)\rightarrow 0$ (see also the green line in Fig.~\ref{fig:bounce_profiles}). In this case $\gamma \simeq 1$ is a good approximation and both the relativistic and non-relativistic approach yield the same result. For general $x$, we provide explicit fitting functions,
\begin{subequations}
\label{eq:semi-analytic}
{\small
\begin{align}\label{eq:fitting_b_global}
\tfrac{R(0;x)}{\mathcal{R}_c}&=\sqrt{1-x} \left(\tfrac{5}{6} + a_1\, x+ a_2\, x^2+ a_3\, x^3+  a_4\, x^4\right)+\tfrac{1}{6}\,, \nonumber\\
b(x)&=(1-x)^{3/2} \left(1 + b_1\, x+ b_2\, x^2+ b_3\, x^3+  b_4\, x^4\right)\,,
\end{align}
}
with numerically determined coefficients
\begin{align}
a_i=\{-0.391,\,0.687,\,-0.854,\,0.363\}\,, \nonumber\\
b_i=\{-0.686,\,2.363,\,-4.058,\,2.686\}\,.
\end{align}
\end{subequations}
Both functions correspond to the black lines in Fig.~\ref{fig:bounce_x_dependence} and provide excellent fits to the numerical results depicted as the black dots.

As in the case of the Coleman bounce, the post-nucleation evolution follows from the analytic continuation of the Euclidean solution to real time. In our setup this implies an $\mathcal{O}(2)\times\mathcal{O}(1,1)$ symmetry for the real-time solution. In the thin-wall regime, the radius of the string at time $t$ is simply $\tilde R(\sqrt{z^2 - t^2})$, where $\tilde R$ is our previous solution. Once $t>z$, the argument becomes imaginary. We analytically continue $\varrho \rightarrow -i\varrho$ in the equations of motion (\ref{bounce:eq}) and solve them numerically along the imaginary axis. Note that this constitutes an initial value problem and no additional shooting is required, since the initial value $R(0)=\sigma \tilde  R_0 / \Delta V$ was already fixed by the original boundary conditions. The corresponding solutions along the imaginary axis are plotted as dashed curves in Fig.~\ref{fig:bounce_profiles}. For this post-nucleation evolution, we have $\gamma \gg 1$, as the bubbles approach the speed of light, invalidating the non-relativistic approach.

\subsection{Local cosmic string}
\label{sec:local_string}
The difficulty in the local case is that beyond the scalar field in \eqref{eq:ansatz_O2_O2}, we also need to make a suitable ansatz for the gauge potential. To that end, we generalize \eqref{ansatz_a} to $A_{\theta}(\varrho, r) =   {n\,a(\varrho, r)}/{g}   $. The action of the local string then becomes
\begin{multline}
\label{S_E_O2O2_field_local}
S_E = 4\pi^2 \int d\varrho\, dr  \varrho \,r\,\Bigg[ \frac{1}{2} (\partial_{\varrho} f)^2+\frac{1}{2}(\partial_{r} f)^2 + V(f) \\
+ \frac{1}{2} \, \frac{n^2}{g^2} \,\frac{1}{r^2}\left[{(\partial_\varrho a)^2+(\partial_r a)^2}\right]+\frac{1}{2} (1-a)^2\,n^2\frac{ f^2}{r^2}\Bigg]\,,
\end{multline}
which in general gives rise to a complicated coupled system of partial differential equations.
As before, we apply the thin-wall approximation by identifying $f(\varrho, r)$ with \eqref{ansatz_O2O2}. For the gauge potential $a$ we use the ansatz in \eqref{eq:ansatz_a_thin_wall} where we replace $R \to R(\varrho)$. We note that strictly speaking this is only a solution of the equations of motion if $\dot R^2,\, R \,\ddot R \ll 1$, which restricts the applicability of this ansatz to a regime where $\gamma \simeq 1$. However, for very large bubble configurations, i.e.\ long after the phase transition, we expect the magnetic contribution to the action to be suppressed due to the additional factor $1/r^2$ in the second row of \eqref{S_E_O2O2_field_local}, making the relativistic regime potentially accessible again within our framework. 
In any event, with these choices the $r$-integration in \eqref{S_E_O2O2_field_local} can  be performed explicitly and we obtain

\begin{multline}
\label{S_thinwall_stringbounce_local}
S_E = 2 \pi \int \varrho\, d\varrho \Bigl[
- \underbrace{\pi R^2(\varrho)\,\Delta V +   \frac{2 \pi\,n^2}{g^2} \frac{1}{R(\varrho)^2}}_{\text{interior}} \\
+ \underbrace{2\pi \,R(\varrho)\sigma \,\gamma^{-1}(\varrho)}_{\text{wall}}
\Bigr]\,,
\end{multline}%
where we kept the $\gamma$ factor in the wall contribution to ensure a relativistic description of the wall sector.
As expected for a local cosmic string there is no contribution from the exterior region with $r>R + \Delta R$. After the rescaling \eqref{eq:rescale}, the action takes the simple form
\beq\label{eq:S_E_local}
S_E =  \frac{4 \pi^2 \sigma^4}{\Delta V^3 } \int d \tilde \varrho \, \tilde \varrho \left[ - \frac{1}{2}\tilde R^2  + \frac{\tilde R}{\gamma} + \left(\frac{3}{8}\right)^3  \frac{y}{\tilde R^2} \right]\,.
\eeq
Thus the bounce dynamics for the local string is determined entirely by the parameter $y$ defined in \eqref{def:y}. It controls the size of the last term, which corresponds to the contribution from the gauge field. We therefore expect to reproduce the $O(4)$ result in the limit $y \to 0$ when the gauge field makes a negligible contribution to the action.

Varying \eqref{eq:S_E_local} with respect to $\tilde R$ yields
\begin{align}
 \ddot{\tilde  R} + \gamma^{-3}\Bigg[1 + y\, \left(\frac{3}{8}\right)^3 \frac{2}{\tilde R^4}\Bigg] + \gamma^{-2}\Bigg[ \frac{ \dot {\tilde  R}}{\tilde \varrho} - \frac{1}{\tilde R}  \Bigg] = 0 \,,
 \label{bounce:eq_local}
\end{align}
where the boundary conditions in \eqref{boundary_cond_string} still apply. 
Solving this equation follows the same steps as in the global string case. The small-$\varrho$ expansion in Eq.~\eqref{small-rho} yields a quadratic term with coefficient $\tilde R_2 =-(1+ y(3/8)^3 2/\tilde R_{0}^4 - 1/\tilde R_{0})/2 $. We then employ a shooting algorithm to obtain the bounce profiles shown in Fig.~\ref{fig:bounce_profiles} (right panel). As in the local case, $R(\varrho)$ approaches the metastable string profile (with $\gamma \simeq 1$) as $y \to 1$. The late ($t>z$) real-time solution can be extracted from the dotted solutions, which are obtained by rotating $\varrho$ into the complex plane through $\varrho \to i \varrho$. 

As before, for general $y$, we provide explicit fitting functions,
\begin{subequations}
\label{eq:semi-analytic_local}
{\small
\begin{align}\label{eq:fitting_b_local}
\tfrac{R(0;y)}{\mathcal{R}_c}&=\frac{\left(1-y\right)^{a_0}}{(1+\sqrt{y})^{3/2}} \left(\tfrac{3}{4} + a_1\, y+ a_2\, y^2+ a_3\, y^3+  a_4\, y^4\right)+\tfrac{1}{4}\,, \nonumber\\
b(y)&=\left(\frac{1-y}{1+y^{b_0}}\right)^{3/2} \left(1 + b_1\, y+ b_2\, y^2+ b_3\, y^3+  b_4\, y^4\right)\,,
\end{align}
}
with numerically determined coefficients
\begin{align}
a_i&=\{0.445,\, 1.286,\,-3.419,\,4.643,\,-2.401\}\,, \nonumber\\
b_i&=\{0.379,\,-0.616,\,1.956,\,-3.191,\,2.43\}\,.
\end{align}
\end{subequations}
Both functions correspond to the blue curves in Fig.~\ref{fig:bounce_x_dependence} and again provide excellent fits to the numerical results. Unlike the global case, we do not present a fully relativistic treatment of the gauge field $a(\varrho, r)$. Consequently, these expressions are accurate only in the limit $y \to 1$ where $\gamma \simeq 1$ (blue solid). For smaller values of $y$ (blue dashed), deviations from our results are expected; by analogy with the global string case, these could reach several tens of percent.\footnote{We still improve over~\cite{Lee:2013zca} by including the $\gamma$ factor in the ansatz for $f(\varrho,r)$, which ensures that \textit{both} limits $y \to 0$ and $y \to 1$ are correctly recovered.}  While this level of accuracy is sufficient for the applications considered in this work, a more precise treatment is left to future studies. 

We emphasize once more that, in both the global and local string cases, the bounce dynamics in the thin-wall limit is fully determined by a single dimensionless parameter ($x$ or $y$, respectively). This dependence enables simple semi-analytic parameterizations of the dimensionless bounce action, $b = B/B_0$. In particular, we find that the $O(2)\times O(2)$ action is always smaller than its $O(4)$ counterpart, i.e.\ $b \leq 1$ for $0 \leq x, y \leq 1$. Phenomenological implications of this result for vacuum tunneling will be discussed in the next section.

\section{Phenomenology}\label{sec:pheno}

In this section we discuss phenomenological implications of the string-induced vacuum decay. We start by comparing the nucleation efficiency of the two decay channels through $O(4)$ and $O(2)\times O(2)$ bubbles in Sec.~\ref{sec:comparison}. This allows us to formulate the condition for the second channel to be dominant. We then discuss the GW production, focusing on modifications to the standard scenario in Sec.~\ref{sec:quadrupole} and \ref{sec:GW}. Our estimates are based on the standard quadrupole approximation for gravitational radiation
\beq
P_{\rm GW} \approx \frac{G}{5} \left(\dddot Q_{ij}\right)^2\,.
\label{quadrupole_formula}
\eeq
Finally, in Sec.~\ref{sec:pheno_model}, we discuss a concrete two-field model, which also takes into account the formation of the metastable cosmic strings.

\subsection{Comparison of nucleation efficiency}
\label{sec:comparison}

The nucleation rate per unit volume for $O(4)$ symmetric  Coleman bubbles can be estimated as~\cite{Coleman:1977py,Callan:1977pt}
\beq\label{eq:Gamma_Coleman}
\frac{\Gamma_{0}}{V} \simeq \left[ \left(\frac{B_0}{2\pi}\right)^2 \frac{1}{\mathcal{R}_c^4} \right] e^{-B_0} \quad  \left[O(4) \right] \,.
\eeq
Here, the bounce action $B_0$ controls the exponential and is defined in \eqref{eq:bounce_action_normalized}. The prefactor, on the other hand, is estimated by a simple dimensional analysis in term of the critical radius $\mathcal{R}_c=3\,\sigma/\Delta V$ and includes the contribution from the zero modes corresponding to the four translational symmetries (in time and space) of the bounce (see also \cite{Ai:2020vhx} for a recent approach). Similarly, in the case of the $O(2) \times O(2)$ configuration, the nucleation rate per unit length of the string is given by
\beq
\frac{\Gamma_{s}}{L_s} \simeq \left[ \left(\frac{b\,B_0}{2\pi}\right) \frac{1}{R(\varrho=0)^2} \right] e^{-b\,B_0 }\quad \left[O(2) \times O(2)\right]\,,
\eeq
where $R(\varrho=0) \sim \mathcal{R}_c$ is the characteristic size of the string-induced bounce at the moment of nucleation and $b(\cdot)$ is given by either \eqref{eq:fitting_b_global} or \eqref{eq:fitting_b_local} in the case of a global or local string, respectively. Now there are only two translational symmetries (along the string axis and in time), reducing the power of the factor $b\,B_0/(2\pi)$ by one unit. To compare the two probabilities one needs to make an assumption about the total length $L_s$ of strings  in a given volume $V$. For instance, if there are approximately $N_s$ Hubble-sized strings per Hubble patch, then $l_s \equiv L_s/V \sim N_s H^2$ and consequently
\beq\label{eq:Gamma_string-induced}
\frac{\Gamma_s}{V} = \frac{\Gamma_s}{L_s} l_s = \left[ \left(\frac{b\,B_0}{2\pi}\right) \frac{N_s H^2}{\mathcal{R}_c^2} \right] e^{-b\, B_0}.
\eeq
One of the central goals of this work is to determine the regime where string-induced decay replaces the standard decay mediated by $O(4)$ bubbles. To that end, we impose the condition that the phase transition completes more rapidly through the string-induced channel. This requires the Hubble rate at percolation to be larger in the string-induced case,
\begin{align}
 H|_{\Gamma_s}  \gg H|_{\Gamma_0}\,.
\end{align}
Assuming a radiation-dominated background and negligible supercooling during the transition, the Hubble rate at percolation satisfies~\cite{Guth:1981uk}
\beq
H^4 \sim \frac{\Gamma}{V}.
\eeq
Substituting the decay rates from Eqs.~\eqref{eq:Gamma_string-induced} and \eqref{eq:Gamma_Coleman}, we find that the string-induced channel dominates when
\beq
e^{ (2b-1)\,B_0} \frac{1}{N_s^2 b^2} \ll 1\,.
\eeq
Equivalently, assuming $N_s$ of order unity, for $b \lesssim 1/2$, string-induced vacuum decay proceeds more efficiently than the conventional bubble nucleation, while for $b \gtrsim 1/2$ the standard Coleman mechanism dominates. Using our numerical results from Eqs.~\eqref{eq:semi-analytic} and~\eqref{eq:semi-analytic_local}, we find that this translates into the approximate parameter conditions
\begin{align}\label{eq:x_y}
0.4\lesssim x \leq 1\,, \quad \text{and}\quad 0.1 \lesssim y \leq 1\,,
\end{align}
for the global and local string-induced decay channels, respectively, to dominate. We note that the lower bound on $y$ might be sensitive to the simplifying assumption we made about the gauge field profile (see Eq.~\eqref{eq:ansatz_a_thin_wall}), although we do not expect a more precise treatment to change it significantly. In any case, using the definitions of $x$ and $y$ in Eqs.~\eqref{def:x} and \eqref{def:y}, these inequalities correspond to simple constraints on the underlying model parameter space and are the main result of this work.

\subsection{Quadrupole moment}
\label{sec:quadrupole}
Here we calculate the quadrupole moment $Q_{ij}$ associated with a single $O(2) \times O(2)$ bubble in the case of the global string. To that end, we evaluate the general expression
\beq\label{def:qpm}
Q_{ij} = \int d^3x\, \left( 3 \, x_i x_j - |\mathbf{x}|^2 \delta_{ij}\right) \rho(\varrho,r)
\eeq
for the bubble after nucleation. Here, the energy density of the field is 
\begin{multline}
\rho(\varrho,r)=\frac{1}{2} (\partial_{\varrho} f)^2+\frac{1}{2}(\partial_{r} f)^2 + V(f) + \frac{1}{2} \,n^2\frac{ f^2}{r^2}
\end{multline}
and we recall that $r=\sqrt{x^2+y^2}$ and $\varrho = \sqrt{z^2 - t^2}$.
Due to its cylindrical symmetry, the only nonzero components are the diagonal ones. Furthermore, $Q_{xx} = Q_{yy} = -Q_{zz}/2$, since the tensor is traceless. Thus it is sufficient to compute the $(zz)$ component, which we denote as $Q \equiv Q_{zz} $,
\beq
Q(t) = 4\pi \int_0^{\infty} dz \int_0^{\infty} r\, dr\, (2 z^2 - r^2) \rho(\varrho, r)\,.
\eeq
This expression can be further simplified in the thin-wall limit. Carrying out the radial integration up to a cutoff $r=R_\mathrm{max}$, we obtain
\begin{multline}
Q(t) = 4\pi \int_0^{\infty} dz \Biggl[ n^2 v_f^2 \Bigl( \ln\left[ \frac{R_\mathrm{max}}{R(\varrho)} \right]z^2 + \frac{R^2(\varrho) - R^2_\mathrm{max}}{4} \Bigr) \\
+ \sigma  \left(2 z^2 - R^2(\varrho)\right) R(\varrho)\, \gamma(\varrho) \left(1+[\partial_z R(\varrho)]^2\right) \\
- \Delta V \left(  R^2(\varrho) z^2 - \frac{R^4(\varrho)}{4} \right) \Biggr]\,,
\end{multline}
where, as in our previous calculations, the radial integration is decomposed into interior, wall, and exterior contributions. It is straightforward to verify that $Q(t)$ vanishes for a spherical bubble, $R(\varrho) = \sqrt{\mathcal{R}_c^2 - \varrho^2}$. Conversely, for a static string configuration with $R(\varrho)=\rm const$ the quadrupole moment diverges. In the context of GW production, the relevant quantity is the third time derivative of $Q$. Therefore, in what follows, we consider the regularized quadrupole moment $Q^\mathrm{(reg)}$, where we subtracted the static string contribution.

\begin{figure}[t]
    \centering
    \includegraphics[width=0.47\textwidth]{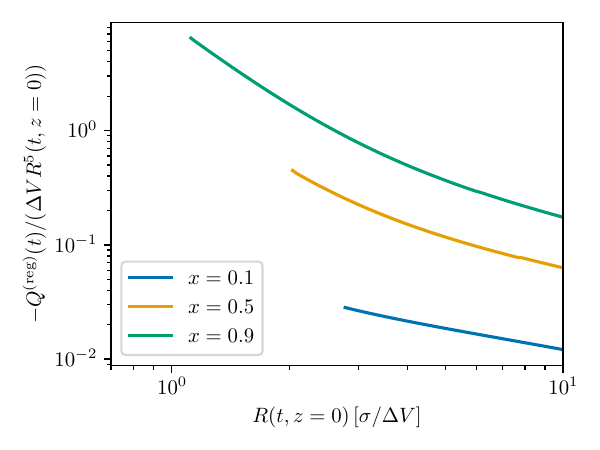}
    \caption{The regularized quadrupole moment $Q^\mathrm{(reg)}$ in units of $\Delta V R^5(t)$, as a function of the radius $R(t)$ for a single $O(2) \times O(2)$ bubble. The same quantity is vanishing for spherical bubbles nucleated in the absence of a cosmic string. 
    }
    \label{fig:quadrupole}
\end{figure}

We compute the quadrupole moment using the numerical solutions for $R(\varrho)$. We find that the wall and the interior contributions are of the same order. The exterior contribution, even if comparable to the other two at $t=0$, grows slower and, thus, is suppressed at late times. All three contributions add up to a negative value that grows approximately as a power-law at late times
\begin{align}\label{eq:power-law}
Q^\mathrm{(reg)}(t) \sim (R(t)/R(0))^{\beta_0}\, Q^\mathrm{(reg)}(0)\,,
\end{align}
where we defined $R(t)=R(\varrho)|_{z=0}$, and numerically inferred the exponent $\beta_0\approx 4$.
 
The fact that the growth is slower compared to $R^5(t)$ reflects that the configuration is more spherical at late times compared to the `naive geometric' scaling of the initial configuration, 
\beq\label{eq:scaling}
R(t,z) = \lambda(t) R(0,z \lambda^{-1}(t))\,,
\eeq
for which $Q^\mathrm{(reg)}(t) = \lambda^5(t) Q^\mathrm{(reg)}(0)$. Evaluating \eqref{eq:scaling} for $z=0$ yields $\lambda(t)=  R(t)/R(t)|_{t=0}$, which indeed implies the above `naive' power law.

In Fig.~\ref{fig:quadrupole}, we plot the evolution of $Q^\mathrm{(reg)}(t)$ for different values of $x$. 
We display $Q^\mathrm{(reg)}(t)$ in units of the time-dependent function $\Delta V R^5(t)$ for two reasons: First, this ratio is dimensionless and depends only on $x$. Second, $\Delta V R^5(t)$ is the naive estimate for the quadrupole moment of two colliding bubbles of size $R$. As can be seen, for $x\simeq 1$ the ratio is initially of order one. In other words, a single $O(2) \times O(2)$ bubble, due to its non-spherical geometry, can have a quadrupole moment as large as that of two colliding Coleman bubbles of the same size. This observation is the basis for our GW analysis in the next section. On the other hand, for $x \to 0$, the ratio is suppressed as the bubble approaches a spherical geometry with vanishing quadrupole moment.

While we have focused on the global cosmic string in this section, we expect qualitatively similar results to hold for the local string case. The main reason is that for large enough bubbles the evolution equations in \eqref{bounce:eq} and \eqref{bounce:eq_local} become indistinguishable.

\subsection{Gravitational waves}
\label{sec:GW}

\begin{figure}[t]
    \centering
    \includegraphics[width=0.47\textwidth]{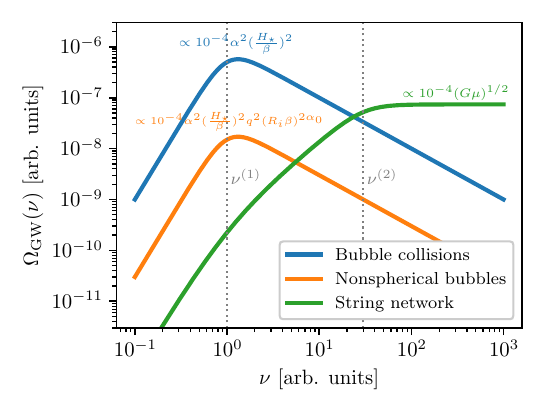}
    \caption{A schematic illustration of the three main sources of GWs associated with the string-induced vacuum decay of a network of local cosmic strings: bubble collisions (blue), expansion of individual non-spherical bubbles (orange), and the loop dynamics in the (transient) string network (green). The first two are peaked around the frequency $\nu^{(1)}$ set by the bubble size at the time of collision (for $\beta/H_{\star} \sim 1$), while the third one is a flat spectrum, suppressed below $\nu^{(2)} > \nu^{(1)}$.}
    \label{fig:schematicGW}
\end{figure}

We now turn to the generation of GWs associated with a string-induced phase transition. We consider three different types of sources. As is typical for first-order phase transitions, the main GW source arises from bubble collisions, which we briefly review. In our scenario, the bubbles are nucleated with a non-spherical geometry. Having computed their quadrupole moment, we estimate the gravitational radiation emitted during the expansion of a single bubble. This is a new type of contribution, which does not exist for standard spherical bubbles.
We then discuss the GW spectrum generated by the dynamics of a transient cosmic-string network. Unlike standard (stable) strings, the network in this case exists only up to the time of the transition, leading to a suppression of the spectrum at low frequencies. The three types of GW sources are summarized schematically in Fig.~\ref{fig:schematicGW}.

GWs from a domain-wall driven phase transition were studied in~\cite{Blasi:2023rqi} by means of hydrodynamic simulations, ignoring the non-sphericity of the bubbles nucleated on domain walls. It was found that the correlation length of the domain wall network can imprint itself in the spectrum of GWs from bubble collisions, enhancing it at low frequencies (see also~\cite{Agrawal:2022hnf}). While a similar effect is also expected to be relevant for string-induced transitions, we do not study it in this work. 

\subsubsection{Bubble collisions}

The peak frequency (at emission) of the gravitational radiation from a phase transition is given by $\nu_{\star} \sim \beta$. Here $1/\beta$ is the duration of the transition which also sets the scale of the initial bubble separation. Redshifting to today and using entropy conservation we find
\beq
\label{eq_freq_GW_PT}
\nu_{\star,0} = \beta  \frac{a_{\star}}{a_0} \approx \beta  \frac{T_0}{T_{\star}}  \sim \frac{\beta}{H_{\star}}  \frac{T_{\star} T_{0}}{M_{\rm Pl}} \sim 10^{-8}\, \mathrm{Hz} \frac{\beta}{H_{\star}} \frac{T_{\star}}{\rm GeV}\,,
\eeq
where a subscript $\star$ denotes evaluation at the time of the phase transition (or bubble collision equivalently), e.g.\ $a_\star=a(t_\star)$, and we ignored for simplicity the change in the number of relativistic degrees of freedom with temperature. Moreover, we assumed that the transition occurred when the universe was dominated by radiation. As for the energy emitted in the form of gravitational radiation, we first estimate the quadrupole moment of a source that consists of two colliding bubbles.
The typical volume of such a configuration is $\sim R^3$, where $R$ is the bubble radius at collision time. The energy density is $\Delta V$ (the interior and wall contributions to the quadrupole moment are of the same order). Here we ignore the interaction of the field with the background radiation, which can generate sound wave dynamics. Due to \eqref{def:qpm}, one expects $Q(R) \sim R^5 \times \Delta V$. The bubbles in vacuum expand approximately at the speed of light, $\dot R \approx 1$, and collide when $R \approx \beta^{-1}$. The power in gravitational radiation is then $P_{\rm GW} \sim G (R^2 \Delta V)^2$ according to Eq.~(\ref{quadrupole_formula}), and the total energy emitted can be estimated as
\beq
E_{\rm GW} = \int P_{\rm GW} \, dt \sim G \beta^{-5} \Delta V^2.
\eeq
As a result, the peak energy density in GWs is $\rho_{\rm GW} \sim G \beta^{-2} \Delta V^2$ and scales as radiation afterwards. The peak fractional energy density in GWs today can be expressed in terms of the radiation density as
\beq
\label{eq_Omega_GW_PT_def}
\Omega_{\rm GW, 0}^{(bc)} = \frac{\rho_{\rm GW}}{\rho_{\rm rad}} \Omega_{\rm rad, 0} \,.
\eeq
Introducing the strength of the phase transition $\alpha = \Delta V/\rho_{\mathrm{rad},\star}$, we estimate
\beq
\label{eq_Omega_GW_PT}
\Omega_{\rm GW, 0}^{(bc)}   \sim  10^{-4} G  \frac{\alpha^2}{\beta^2} \rho_{\rm rad, \star} \sim 10^{-4} \alpha^2 \left( \frac{H_{\star}}{\beta} \right)^2\,.
\eeq
In the absence of significant supercooling, we have $\alpha \ll 1$. Eq.~(\ref{eq_freq_GW_PT}) and (\ref{eq_Omega_GW_PT}) agree well with the standard estimates for the GW frequency and amplitude, respectively (see e.g.~\cite{Caprini:2015zlo}). The spectrum is schematically illustrated in Fig.~\ref{fig:schematicGW} as the blue line. We stress that these order unity estimates are equally valid for $O(4)$ and $O(2) \times O(2)$ bubbles. In the latter case we simply identify $R = R(t_\star)$.

\subsubsection{Non-spherical bubbles}

In the case of (local and global) string-induced decay,
bubbles are not spherical and therefore carry a nonvanishing quadrupole moment. As shown in Sec.~\ref{sec:quadrupole}, this quadrupole moment grows with time until the bubbles collide, providing an additional source of gravitational radiation.

The characteristic frequency is given by the inverse lifetime of the source and thus we expect it again to be given by (\ref{eq_freq_GW_PT}).
To estimate the energy density in GWs, we use the numerically extracted power-law in \eqref{eq:power-law} to extrapolate to very late times.

We find that
\beq
Q(t) = \Delta V R(t)^5 \Bigl[ \frac{R(t)}{\sigma/\Delta V} \Bigr]^{-\alpha_0} q(\cdot),
\eeq
where $R(t)= R(\varrho)|_{z=0} $ and $\alpha_0=5-\beta_0 \approx 1$, using our numerical result from Sec.~\ref{sec:quadrupole}. The function $q(\cdot)$ encodes the initial nonsphericity and is order one for $x,y \simeq 1$. Computing the third derivative and inserting into the formula for the gravitational radiation one finds that
\beq
\Omega_{\rm GW, 0}^{(nb)} \sim \Omega_{\rm GW, 0}^{(bc)}  q^2\left[ \frac{\sigma/\Delta V}{\beta^{-1}} \right]^{2\alpha_0} \sim \Omega_{\rm GW, 0}^{(bc)} q^2 ( R_i\beta )^{2\alpha_0} \,,
\eeq
where $R_i=R(\varrho)|_{\varrho=0}$. In other words, the GW contribution from the expansion of individual bubbles is suppressed by $R_i/\beta^{-1}$, which is the ratio of the initial and final sizes of the bubbles. If the bubbles collide after growing by a large amount from the time of nucleation, which is usually the case, the suppression is strong and the sourcing of GWs due to the non-spherical bubble geometry is a comparably small effect. However, if the percolation is extremely rapid with $1/\beta \sim R_i$, or if $x \rightarrow 1$, the non-sphericity can make a sizeable contribution. This new type of GW source is schematically illustrated in Fig.~\ref{fig:schematicGW} as the orange line.

\subsubsection{String network}

In contrast to the previous two sources, the cosmic string network radiates GWs at a wide range of frequencies. 

The GW background from a local string network can be expressed as (see e.g.~\cite{Gouttenoire:2019kij, Servant:2023tua})
\beq
\label{eq:GW_spectrum}
\begin{split}
&\Omega_{\mathrm{GW}, 0}(\nu) = G\mu_s^2 \frac{1}{\rho_0} \sum_k \Gamma^{(k)}  \\&\times \Bigl( \frac{2k}{\nu} \Bigr) \int_{t_{\rm SSB}}^{t_0} dt\left( \frac{a(t)}{a_0} \right)^5 n_{\rm loop} \Bigl( \frac{2k}{\nu} \frac{a(t)}{a_0}, t\Bigr)\,,
\end{split}
\eeq
where $\rho_0$ is the total energy density today, i.e.\ $\rho_0= \rho_\mathrm{tot}(t_0)$, and $\Gamma^{(k)}=\Gamma k^{-n}/\sum_{p=1}^{\infty} p^{-n}$ is the fractional decay rate into the $k$-th harmonic with $n=4/3$ assuming cusps dominating the small-scale structure of the string~\cite{Vachaspati:1984gt}, and $l_k = 2k/\nu$ is the comoving length scale of the $k$-th harmonic of the string, corresponding to radiation observed at the frequency $\nu$ today. Moreover, $n_{\rm loop}(l, t)$ is the number density of strings of physical length $l$ at time $t$. In the scaling regime it takes the form
\beq
n_{\rm loop}(l, t) = \frac{\mathcal{F}\,C_{\rm eff}}{\alpha_1 (\alpha_1 + \Gamma G \mu_s) } \frac{1}{t_{\rm ini}^4(l,t)}  \left( \frac{a(t_{\rm ini}(l,t))}{a(t)} \right)^3
\eeq
for $t_{\rm SSB}<t_{\rm ini}(l,t) < t<t_0$ and is zero otherwise. Here $t_{\rm ini}(l,t)$ is the time of formation of the strings that, at time $t$, have physical size $l$. It is given by
\beq
t_{\rm ini}(l,t) = \frac{l+\Gamma G\mu_s t}{\alpha_1 + G \Gamma \mu_s}\,,
\eeq
where $\alpha_1\approx 0.1$, $\mathcal{F}\approx 0.1$ and $C_{\rm eff}\approx 5.3 (0.4)$ during the radiation (matter)-dominated era.  The later formula follows from
\beq
l(t) = l(t_{\rm ini}) - \Gamma G \mu_s (t-t_{\rm ini})
\eeq
for $t>t_{\rm ini}$ with the identification $l(t_{\rm ini}) = \alpha_1\, t_{\rm ini}$.

In order to implement the vacuum decay of the string network, we introduce a cutoff at $t=t_\star$ in the time integration in \eqref{eq:GW_spectrum}. This is equivalent to setting $n_{\rm loop}(l,t)=0$ for $t>t_{\star}$. The standard result is then recovered for the choice $t_\star=t_0$.  
Assuming that the GW emission takes place deep inside the radiation era, $a(t) = c\sqrt{t}$, and focusing on the first harmonic with $k=1$, the integral in~(\ref{eq:GW_spectrum}) can be performed analytically,  
\beq
\Omega_{\mathrm{GW}, 0}(\nu)  = \Omega_{\mathrm{GW}, 0}^\mathrm{(stable)}\times \mathcal{K}(\nu),
\eeq
where 
\beq
\begin{split}
\Omega_{\rm GW,0}^\mathrm{(stable)} &= \frac{\Gamma^{(1)} G \mu_s^2 }{\rho_0} \frac{4\mathcal{F}C_{\rm eff} c^4}{3\alpha } \frac{(\alpha + \Gamma G \mu_s)^{3/2}}{{(\Gamma G \mu_s )}^{3/2}}\\
& \sim 4 \times 10^{-9} \sqrt{\frac{G\mu_s}{10^{-7}} }\,.
\end{split}
\eeq
matches the usual expression for the flat GW spectrum emitted from a stable local string network. To arrive at the second line, we set $\Gamma=50$ and $c = 1.1(H_0^2 \Omega_{\mathrm{rad}, 0})^{1/4}$. The suppression of the spectrum caused by the vacuum decay is then captured by
\beq
\mathcal{K}(\nu) \approx \, \Bigl( \frac{\nu}{\nu + \nu^{(2)}}\Big)^{3/2} - \Bigl( \frac{\nu^{(1)}}{\nu^{(1)} + \nu^{(2)}}\Big)^{3/2}\,.
\eeq
where we defined
\beq
\nu^{(1)} = \frac{2}{\alpha_1} \frac{1}{t_{\star}}\frac{a_{\star}}{a_0} \sim H_{\star}  \frac{T_0}{T_{\star}} \sim 10^{-8} \,\mathrm{Hz}\, \frac{T_{\star}}{\rm GeV}\,,
\eeq
and
\beq
\label{eq:nu2}
\nu^{(2)} =  \frac{2}{\Gamma G \mu_s} \frac{1}{t_{\star}}\frac{a_{\star}}{a_0} \sim 10^{-3}\,\mathrm{Hz}\, \frac{T_{\star}}{\rm GeV} \frac{10^{-7}}{G\mu_s}\,,
\eeq
where $\nu^{(2)} \gg \nu^{(1)}$ for sufficiently sub-Planckian string tensions. There are three important frequency ranges: for $\nu \gg \nu^{(2)}$ the spectrum is approximately flat since $\mathcal{K}(\nu) \approx 1$, and one recovers the result for a stable network. In the intermediate range of frequencies $\nu^{(1)} <\nu< \nu^{(2)}$ one observes a tilted spectrum $\propto \mathcal{K}(\nu) \approx (\nu/\nu^{(2)})^{3/2}$. Around $\nu\approx \nu^{(1)}$ there is a sharper drop in the spectrum as $\mathcal{K}(\nu)\rightarrow 0$. Here one expects a causal $\propto \nu^3$ tail, as for any GW source that is localized in time. Such a spectrum is shown in Fig.~\ref{fig:schematicGW} as the green line.

Including the higher harmonics does not affect much the growing part of the spectrum, but softens the transition to the flat region and increases its amplitude  by approximately a factor three.

Overall, the effect of the phase transition is to suppress the GW spectrum for frequencies below $\nu^{(2)}$ according to (\ref{eq:nu2}), which is a consequence of limiting the time integration in~(\ref{eq:GW_spectrum}) to times before $t_{*}$. Such a scenario can have interesting implications for bounds on local cosmic strings from PTAs, as well as the CMB. While stable local strings with $G\mu_s \gtrsim 10^{-10}$ are currently excluded by NANOGrav 15-year data~\cite{NANOGrav:2023hvm}, below we demonstrate how these bounds can be evaded in the presence of a symmetry-restoring phase transition. 

\begin{figure*}[t]
    \centering
    \includegraphics[width=0.47\textwidth]{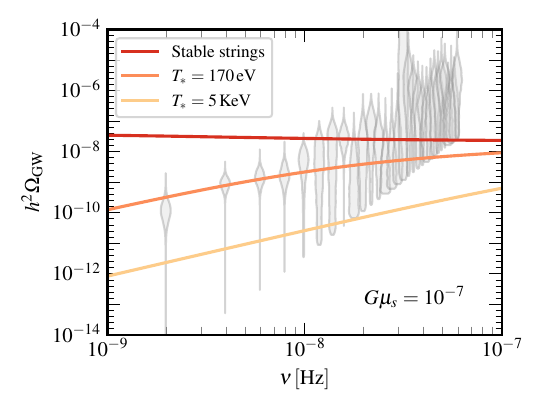}
    \includegraphics[width=0.47\textwidth]{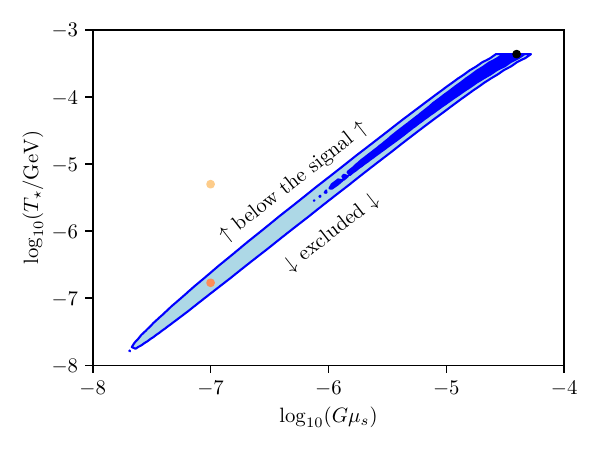}
    \caption{\textbf{Left:} GW spectra from a metastable local string network for different values of the temperature $T_\mathrm{*}$ of a symmetry restoring phase transition. The violin shapes indicate the uncertainties of the recent NANOGrav 15-year observations~\cite{NANOGrav:2023gor}. 
    The chosen string tension, $G\, \mu_s=10^{-7}$ is typical of a GUT-scale transition and would be excluded for a stable network (red). If the phase transition occurs sufficiently early (yellow), the low-frequency part of the spectrum is suppressed, allowing the bound to be evaded. For an intermediate value, $T_\mathrm{*} = 170 \,\mathrm{eV}$ (orange), corresponding to a post-BBN phase transition, the predicted and observed signals are of comparable magnitude. \textbf{Right:} Reconstructed $1\sigma$ and $2\sigma$ confidence intervals from a frequentist analysis of the NANOGrav 15-year data. The values of $G\, \mu_s$ and $T_\mathrm{*}$ corresponding to the best fit and to the orange and yellow lines in the left panel are shown as black, orange and yellow dots, respectively. Overall, our scenario provides an excellent fit to NANOGrav data with a best fit corresponding to $\chi^2/\text{dof} \approx 0.5$.}
    \label{fig:nanograv}
\end{figure*}

To compare the GW signal with data, we compute the integral in (\ref{eq:GW_spectrum}) numerically, using the full $\Lambda$CDM expansion history with Planck $2018$ parameters~\cite{Planck:2018vyg}, taking into account changes in the number of relativistic degrees of freedom~\cite{Husdal:2016haj}. This slightly modifies the low-frequency behavior compared to the radiation-dominated approximation discussed above. We include the first $10^4$ harmonics in the sum in \eqref{eq:GW_spectrum} to ensure good convergence. As demonstrated in the left panel of Fig.~\ref{fig:nanograv} for the fiducial choice $G \mu_s = 10^{-7}$, if the transition occurs at temperatures \textit{above} $T_{*} \sim 100 \,\rm eV$, the spectrum is suppressed around the frequencies probed by PTAs and the corresponding bounds are evaded (yellow line). However, there is the even more exciting possibility that for temperatures of order of $T_{*} \sim 100 \,\rm eV$, we explain the observed signal rather than just evading the bounds (orange line).

In the right panel of Fig.~\ref{fig:nanograv} we show the results of a frequentist analysis of the NANOGrav 15-year data, following the method in~\cite{Winkler:2024olr}. We approximate the posterior distributions in each frequency bin, i.e.~the gray `violins', as piecewise Gaussian in $\log_{10}\Omega_{\mathrm{GW}}$, using the median $\log_{10}\bar{\Omega}_{\mathrm{GW}, i}$ as the central value and separate variances $\Delta \log_{10}{\Omega}_{\mathrm{GW}, i}$ for upward and downward fluctuations. For given values of $\mu_s$ and $T_{\star}$, we compute $\chi^2$ as 
\beq
\chi^2 = \sum_{i \in \rm bins} \frac{ (\log_{10}\Omega_{\rm GW}(\nu_i) - \log_{10}\bar{\Omega}_{\mathrm{GW}, i})^2 }{(\Delta\log_{10}{\Omega}_{\mathrm{GW}, i})^2},
\eeq
summing only over the first $14$ bins to eliminate the high-frequency noise. The best-fit parameters (shown as the black dot) correspond to $\chi^2_{\rm min}$, and the $1\sigma$ ($2\sigma$) contours are identified with $\chi^2 - \chi^2_{\rm min} = 2.3\, (6.2)$.  The region above the contours evades the PTA bounds. Inside the contours the predicted GW signal itself provides a good fit to data. In particular, we find $\chi^2/\text{dof} \approx 0.5$ for the best fit (shown as black dot) and $\chi^2/\text{dof} \approx 0.8$ for parameters corresponding to the orange line in the left panel (shown as orange dot in the right panel). The number of degrees of freedom (dof) is $12 = 14-2$, where we subtract the number of model parameters. In contrast, the best fit value for the case of stable strings is $\log_{10}(G\mu_s) \approx -10.2$ and has $\chi^2/\text{dof} \approx 1.7$. 
An abrupt increase of $\chi^2$ for $G\mu_s>10^{-4}$ is due to the pulsars becoming sensitive to frequencies around $\nu^{(1)}$, where the spectrum has a sharp drop. This abrupt behavior is an artifact of describing the symmetry-restoring phase transition as an instantaneous process.


In summary, PTA data prefers a symmetry-restoring phase transition occurring between BBN and recombination. Such a transition is not constrained by the CMB provided the latent heat is small. This is achieved when the two minima of the potential are nearly degenerate, so that the vacuum energy difference is small compared to the radiation energy density. Moreover, the string network carries only a negligible fraction $\sim G\mu_s$ of the total energy density, so its decay does not lead to observable entropy production or distort the CMB. Consequently, such a late  transition can occur without conflicting with any cosmological bounds.

This mechanism of alleviating the bounds on local cosmic strings is similar to other metastable cosmic string scenarios. For example, in the scenario proposed in \cite{Vilenkin:1982hm}, strings can attach to magnetic monopoles and decay via monopole–antimonopole annihilation, suppressing the GW background~\cite{Martin:1996ea,Martin:1996cp,Leblond:2009fq,Buchmuller:2019gfy}.

For completeness, we briefly comment on networks of global strings. In contrast to local strings, global strings radiate predominantly into scalar waves corresponding to the massless Goldstone bosons of the spontaneously broken $U(1)$ symmetry. The resulting GW background is suppressed in amplitude relative to the local case and exhibits a tilted spectrum that decreases with frequency during the radiation era, rather than remaining flat~\cite{Chang:2019mza}. A more comprehensive analysis of the GW signatures from both local and global string networks in the context of string-induced vacuum decay is left for future work.

\subsection{A concrete realization}
\label{sec:pheno_model}

So far we have not included the creation of the cosmic strings in our description and instead simply assumed their presence. The aim of this section is to provide a  two-field model that extends our minimal example from Sec.~\ref{sec:minimal_setup} and accounts for the creation and subsequent (quantum) decay of the cosmic string.  
To that end, consider the following action of two coupled scalar fields,
\beq
 S=\int d^4 x \Bigl[ |\partial \phi|^2 +\frac{1}{2} (\partial \chi)^2 -  V(\phi, \chi)\Bigr] + \ldots \,,
 \nonumber
\eeq
where the ellipses stand for additional gauge couplings in the case of the local strings or higher terms in an EFT expansion. Here, we introduced in addition to the complex tunneling field $\phi$ from before a real scalar $\chi$.  The 2-field potential is $V(\phi, \chi) = V_{\phi}(\phi) + V_{\chi}(\chi)  + V_{\rm int}(\phi, \chi)$, where $V(\phi)$ is obtained from Eq.~\eqref{eq:V} in our minimal model through the replacement $\mu^2 \to -\mu^2$,
\begin{align}
 V_{\phi}(\phi) = -\mu^2 |\phi|^2 - \lambda |\phi|^4 + \lambda_6 |\phi|^6 \,.
\end{align}

\begin{figure}[t]
    \centering
    \includegraphics[width=0.47\textwidth]{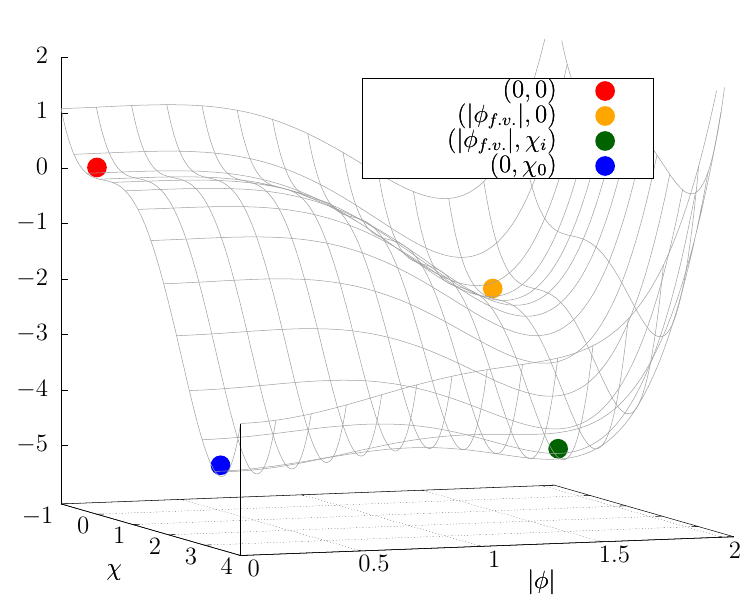}
    \caption{The two-field potential $V(\phi, \chi)$, defined in~(\ref{define_Vfx}), for some benchmark parameters. All dimensionful quantities are in units of $ \lambda/(2\sqrt{\lambda_6})$. The field starts at $(0,0)$ (red point) for high temperatures. Below a critical temperature $ T_c$, it rolls down to  $(|\phi_\mathrm{f.v.}|, 0)$ (orange dot), spontaneously breaking the $U(1)$ symmetry and forming cosmic strings via the Kibble mechanism. Subsequently, when the Hubble friction is small enough, the field settles into the metastable local minimum at $(|\phi_\mathrm{f.v.}|, \chi_i)$ (green dot). The corresponding vacuum state is metastable and the field eventually tunnels to  $(0, \chi_0)$ (blue dot) through the string-induced decay studied in this work. }
    \label{fig:twofieldpotential}
\end{figure}

In addition, the real field $\chi$ comes with the potential
\begin{align}
V_{\chi}(\chi) = \frac{m^2}{2}\chi^2 - \frac{\lambda_3}{3} \chi^3 + \frac{\lambda_{\chi}}{4} \chi^4 \,,
\end{align}
where $m^2>0$, $\lambda_3>0$ and $\lambda_{\chi}>0$. We require that $V_{\chi}$ has a false minimum at $\chi=0$ and a true minimum, which exists at 
\begin{align}
\chi_0 = \frac{\lambda_3}{\lambda_{\phi}} \left( 1 + \sqrt{1 - \frac{2m^2 \lambda_{\phi}}{\lambda_3^2} }\right),
\end{align}
if we impose $0 < m^2 < 4\,\lambda_3^2/(9\lambda_{\phi})$. Finally, the interaction between both field is described by
\begin{align}
V_{\rm int}(\phi, \chi ) = (\xi \, \chi^2 - \eta \, \chi )|\phi|^2 \,,
\label{define_Vfx}
\end{align}
with dimensionless couplings $\xi>0$ and $\eta>0$. The potential is depicted in Fig.~\ref{fig:twofieldpotential} for a set of benchmark values.\footnote{For completeness,  we set $\lambda=1$ and express all dimensionful quantities in units of $M=\lambda/(2\sqrt{\lambda_6})$. The choice $m=0.78 \,M$, $\lambda_3=1.93 \, M$, $\lambda_{\chi}=1$, $\xi=0.11$ and $\eta=0.07\,M$ results in the two-field potential shown in Fig.~\ref{fig:twofieldpotential}. The metastable minimum for this choice is at $\chi_0 = 3.23\mu_0$ and $\phi_i =  1.44\mu_0$, while the true minimum has $\chi_0 = 3.5\,M$ and corresponds to $\mu(\chi_0) = 1.1\,M$.
Note that the distance between the two minima in the $\chi$-direction, $\chi_0- \chi_i$, is an order of magnitude smaller compared to corresponding distance in the direction of the canonically normalized $|\phi|$ field. Hence, tunneling is mainly in the $\phi$-direction.} In short, the effect of the new field $\chi$ is to promote the parameter $\epsilon$ in \eqref{eq:epsilon} to a function 
\begin{align}\label{eq:epsilono}
\epsilon(\chi) = -2 + \frac{\lambda \left(4 \lambda - 3 \lambda_6 v_f^2 \right)}{2\, \lambda_6\, \mu_\mathrm{eff}(\chi)^2}\,,
\end{align}
where we introduced the $\chi$-dependent mass for the $\phi$ field of the form
\beq
\mu_\mathrm{eff}^2(\chi) = -\mu^2 +\xi \, \chi^2 - \eta \,\chi\,.
\eeq
Now, as $\chi$ evolves, we can transition from a regime where the potential supports stable cosmic strings ($\epsilon<0$), making the standard formation mechanism applicable, to a regime where the strings become metastable ($\epsilon>0$). These two limiting regimes correspond to the dashed and solid lines in  Fig.~\ref{fig:V}, respectively. 

To get a better understanding of the intermediate evolution, we consider Fig.~\ref{fig:twofieldpotential}. The different stages of the formation and decay process are marked as the colored points and correspond to the sequence
\begin{equation}
\begin{array}{ccc}
{\color{red}(0,0)}  & \xrightarrow[\text{strings form}]{} & {\color{orange}(|\phi_\mathrm{f.v.}|, 0)} \\
\quad\quad\cancel{\downarrow}\!{\scriptstyle\text{  blocked}}& & \downarrow\!\!{\scriptstyle\text{  roll-over}} \\[+0.3em]
{\color{blue}(0, \chi_0)} & \xleftarrow[\text{string-induced decay}]{} & {\color{darkgreen}(|\phi_\mathrm{f.v.}|, \chi_i)}
\nonumber
\end{array}
\end{equation}

Let us now go through the individual steps in more detail.  

At high temperatures, the gauge symmetry is restored and the field is expected to be at $(|\phi|, \chi) = (0,0)$ (red dot). Spontaneous symmetry breaking occurs near the critical temperature $T_c$, defined through $\mu_\mathrm{eff}^2(0) + c_0\,T_c^2 \simeq 0$~\cite{Dolan:1973qd}, where $c_0$ is an order unity constant that depends on the details of the underlying theory such as the couplings to the thermal bath. 
Due to the positive mass $m^2>0$, the field cannot role (or tunnel) to the global minimum (blue dot) directly but needs to take a `detour' along the $|\phi|$ direction (with  $\chi\simeq 0$) first. Since initially $\mu_\mathrm{eff} < 0$ and thus $\epsilon<0$, strings form via the Kibble mechanism~\cite{Kibble:1976sj, Kibble:1980mv, Vilenkin:1981kz} as $\phi$ rolls down towards $|\phi_\mathrm{f.v.}|=(v_f(\chi)/\sqrt{2})|_{\chi=0}$ (orange dot). Here, we generalized the definition \eqref{eq:vf} to
\begin{align}
v_f^2(\chi)=\frac{2}{3 \lambda_6}\left(\lambda + \sqrt{\lambda^2 -  3 \lambda_6 \,\mu_\mathrm{eff}^2(\chi)}\right)\,.
\end{align}
At this point, the potential along the $\chi$-direction is no longer stabilized by a barrier. This is the case if $V_{\chi}(\chi) + V_{\rm int}(v_f(\chi)/\sqrt{2}, \chi)$ decreases with $\chi$ near $\chi = 0$, which imposes only a mild constraint on the model parameters. We note, however, that if one aims not only to evade the PTA constraints but to explain the observed signal, the potential must be considerably flatter in the $\chi$ direction than in the $|\phi|$ direction. This would delay the roll-over toward the false minimum and thus postpone the onset of tunneling. Whether such a hierarchy of scales can be achieved in a technically natural way consistent with observational bounds we leave for future model-building work. In any case, the field eventually settles into the metastable minimum at $(|\phi|, \chi) = (v_f(\chi_i)/\sqrt{2}, \chi_i)$ (green dot). Requiring the generalized bound in~\eqref{eq:bounds} to hold, i.e.\
\begin{align}
1/4<\frac{\lambda_6 \mu_\mathrm{eff}^2(\chi_i)}{\lambda^2}<1/3\,,
\end{align}
we obtain $\epsilon(\chi_i)>0$. This shows that we have successfully prepared the initial state of a metastable cosmic string as considered in the main analysis of this work. The tunneling probability to the true minimum (blue dot) is then obtained from our previous results when identifying 
\begin{align}
\left\{\mu_\mathrm{eff}(\chi),v_f(\chi),\epsilon(\chi)\right\}|_{\chi=\chi_i \simeq\chi_0} \leftrightarrow \left\{\mu,v_f,\epsilon\right\}\,.
\end{align}

In addition to quantum tunneling, studied in this work, thermal fluctuations can also play a role in the string-induced decay due to the coupling to the thermal bath~\cite{Linde:1981zj}. However, these effects can be suppressed by reducing the value of $|\mu^2_\mathrm{eff}(0)|=\mu^2$, which lowers the temperature at which the first stage of the transition occurs.

\section{Conclusion}
\label{sec:conclusion}

In this work we have studied the false vacuum decay catalyzed by cosmic strings.
In Eq.~(\ref{eq:x_y}), we derived the parameter regime in which bubbles of true vacuum are predominantly nucleated along the string axis, triggering an instability of the string core that propagates outward at nearly the speed of light. As these $O(2) \times O(2)$ bubbles expand and collide, they generically dissipate into radiation, converting space to the symmetric true vacuum state and effectively erasing the catalyzing string. 

This mechanism works for both global and local strings, and we have derived explicit expressions for the corresponding tunneling rates in the thin-wall limit. For the global case, we used a fully relativistic treatment that provides a reliable description of the bubble-wall evolution after the phase transition. For local strings, we relied on a simplified ansatz where the gauge potential adiabatically tracks the radial profile of a corresponding static string. We argued that this approach provides acceptable accuracy for calculating the bounce action as well as describing the post-nucleation dynamics. 
Extending the analysis beyond this approximation will require solving the coupled partial differential equations in both the Euclidean radius $\varrho$ and the polar radius $r$, in the presence of a moving boundary at $r = R(\varrho)$. This task is left to future work.

An especially interesting application of string-induced vacuum decay concerns GWs. First, we identify a new contribution arising from the non-sphericity of the string-induced bubbles. Although this contribution is generally subdominant compared to that from bubble-wall collisions, it can become comparable when percolation proceeds very rapidly. Second, cosmic-string networks are known to generate a stochastic GW background characterized by a broad plateau in frequency space. Recent PTA observations have placed stringent upper limits on the string tension, effectively excluding standard GUT-scale scenarios. The metastability of the cosmic-string vacuum, however, may provide a natural mechanism to evade these bounds, offering an alternative to other decay mechanisms such as string fragmentation via monopole–antimonopole nucleation~\cite{Vilenkin:1982hm,Martin:1996ea,Martin:1996cp,Leblond:2009fq,Buchmuller:2019gfy} (see also~\cite{Kamada:2015iga,Bettoni:2018pbl,Gouttenoire:2019kij} for related scenarios).

Beyond the possibility of evading GW constraints, it is intriguing to consider whether the observed PTA signal could in fact originate from a metastable cosmic-string network. As a first step in this direction, we have shown that a post-BBN vacuum phase transition can generate a spectrum with an amplitude that provides a good fit to NANOGrav data. We have also introduced a minimal two-field model that realizes the spontaneous symmetry breaking responsible for string formation and subsequently induces a delayed symmetry restoration through a string-induced vacuum decay. A more detailed investigation, which compares different metastable string scenarios and uses the data fit to constrain the details of the two-field model, is left for future work.

Finally, within the new early dark energy framework~\cite{Niedermann:2023ssr} it has recently been shown that a supercooled (\textit{hot new early dark energy})~\cite{Niedermann:2021vgd,Niedermann:2021ijp,Garny:2024ums,Garny:2025kqj} or vacuum phase transition (\textit{cold new early dark energy}~\cite{Niedermann:2019olb,Niedermann:2020dwg,Chatrchyan:2024xjj}) occurring after BBN and injecting energy into the dark sector fluid can resolve the Hubble tension, a discrepancy in the observed value of the Hubble constant~\cite{Riess:2021jrx} (for a review see~\cite{CosmoVerseNetwork:2025alb}). It is therefore natural to ask whether the phase transition discussed here is compatible with that framework and could provide a strong enough energy injection into the dark sector to address the tension while also explaining the NANOGrav signal (for an alternative idea see~\cite{Cruz:2023lnq}). In particular, it might be possible to embed the trigger mechanism discussed in~\cite{Niedermann:2021ijp} into the two-field model discussed in Sec.~\ref{sec:pheno_model}.

\begin{acknowledgments}
We would like to thank Edmund Copeland, Joerg Jaeckel, Martin S.~Sloth, and Jes\'us Zavala for useful discussions. 
The work of F.N.\ is supported by VR Starting Grant 2022-03160 of the Swedish Research Council. The work of A.C. was supported by the Swedish Research Council (VR) under grants 2018-03641 and 2019-02337.
A.C.\ and F.N.\ thank the Center for Astrophysics and Cosmology at the University of Iceland for its hospitality.
\end{acknowledgments}

\end{document}